\documentclass[aip,reprint]{revtex4-1}

\usepackage[utf8]{inputenc}

\usepackage{graphicx}

\begin{document}


\title{Over-critical sharp-gradient plasma slab produced by the collision of laser-induced blast-waves in a gas jet; Application to high-energy proton acceleration.}



\author{J.-R. Marquès}
\email[]{jean-raphael.marques@polytechnique.fr}
\affiliation{LULI, CNRS, École Polytechnique, CEA, Sorbonne Université, Institut Polytechnique de Paris, F-91128 Palaiseau Cedex, France}

\author{P. Loiseau}
\affiliation{CEA, DAM, DIF, F-91297 Arpajon Cedex, France}
\author{J. Bonvalet}
\affiliation{CELIA, Université de Bordeaux–CNRS–CEA, Talence 33405, France}
\author{M. Tarisien}
\affiliation{CENBG, CNRS-IN2P3, Université de Bordeaux, 33175 Gradignan Cedex, France}
\author{E. d’Humières}
\affiliation{CELIA, Université de Bordeaux–CNRS–CEA, Talence 33405, France}
\author{J. Domange}
\affiliation{CENBG, CNRS-IN2P3, Université de Bordeaux, 33175 Gradignan Cedex, France}
\author{F. Hannachi}
\affiliation{CENBG, CNRS-IN2P3, Université de Bordeaux, 33175 Gradignan Cedex, France}
\author{L. Lancia}
\affiliation{LULI, CNRS, École Polytechnique, CEA, Sorbonne Université, Institut Polytechnique de Paris, F-91128 Palaiseau Cedex, France}
\author{O. Larroche}
\affiliation{CEA, DAM, DIF, F-91297 Arpajon Cedex, France}
\author{P. Nicolaï}
\affiliation{CELIA, Université de Bordeaux–CNRS–CEA, Talence 33405, France}
\author{P. Puyuelo-Valdes}
\affiliation{CENBG, CNRS-IN2P3, Université de Bordeaux, 33175 Gradignan Cedex, France}
\author{L. Romagnani}
\affiliation{LULI, CNRS, École Polytechnique, CEA, Sorbonne Université, Institut Polytechnique de Paris, F-91128 Palaiseau Cedex, France}
\author{J. Santos}
\affiliation{CELIA, Université de Bordeaux–CNRS–CEA, Talence 33405, France}
\author{V. Tikhonchuk}
\affiliation{ELI-Beamlines, Institute of Physics, Academy of Sciences of the Czech Republic, 18221 Prague, Czech Republic}
\affiliation{CELIA, Université de Bordeaux–CNRS–CEA, Talence 33405, France}

\date{\today}

\begin{abstract}
The generation of thin and high density plasma slabs at high repetition rate is a key issue for ultra-high intensity laser applications. We present a scheme to create such plasma slabs, based on the propagation and collision in a gas jet of two counter-propagating blast waves (BW). Each BW is launched by a sudden and local heating induced by a nanosecond laser beam that propagates along the side of the jet. The resulting cylindrical BW expands perpendicular to the beam. The shock front, bent by the gas jet density gradient, pushes and compresses the plasma toward the jet center. By using two parallel ns laser beams, this scheme enables to tailor independently two opposite sides of the jet, while avoiding the damage risks associated with counterpropagating laser beams. A parametric study is performed using two and three dimensional hydrodynamic, as well as kinetic simulations. The BWs bending combined with the collision in a stagnation regime increases the density by more than 10 times and generates a very thin (down to few microns), near to over-critical plasma slab with a high density contrast ($>$ 100), and a lifetime of a few hundred picoseconds. Two dimensional particle-in-cell simulations are used to study the influence of plasma tailoring on proton acceleration by a high-intensity sub-picosecond laser pulse. Tailoring the plasma not only at the entrance but also the exit side of the ps-pulse enhances the proton beam collimation, increases significantly the number of high energy protons, as well as their maximum energy.
\end{abstract}


\maketitle 

\section{Introduction}\label{Intro}
With the advent of high-repetition rate ultra-high power lasers, new applications have emerged, some of which requiring the use of very dense and thin plasma slabs. Among them we find transient plasma photonic crystals \cite{Lehmann_PRL,Lehmann_PRE} that are considered as a promising solution to overcome the limitations caused by the damage threshold of solid-state optical materials when manipulating the next generation ultra-high-power lasers. Another example is the production of femtosecond MeV electron bunches by single-cycle laser pulses \cite{Faure}, which could be an attractive solution for ultrafast imaging or femtosecond x-ray generation. The  generation of high energy ion beams from laser-induced Collisionless Shock Acceleration (CSA) \cite{Denavit,Silva,Fiuza_2012,Macchi,Grassi} is an additional application of major interest and is the one we will focus on in this paper. Typically these applications require a plasma slab with thickness $<$ 100 $\mu$m, sharp gradients ($<$ 10 $\mu$m), and tunable density ($10^{20}$-$10^{21}$ cm$^{-3}$), and that could be operated at high-repetition rates ($>$ 10 Hz). Such a device is very challenging to produce, and several schemes have been recently proposed/tested, particularly in the context of ion-acceleration by lasers, where the ion beam characteristics strongly depend on the plasma density gradients.

Up to now, laser-produced ion beams have been mostly generated via the target normal sheath acceleration (TNSA) mechanism\cite{Wilks, Macchi}. This mechanism is very robust, but requires solid targets which can be unpractical for applications at high repetition rate and in a debris free environment. Lately, other advanced acceleration schemes have been proposed and studied. One that can be fully explored with current laser intensities ($10^{19}-10^{21}$ W/cm$^2$) is the CSA. It relies on accelerating ions through reflection from a moving shock wave in a plasma of electron density $n_e$ close to the critical density $n_c$ for the driving laser ($n_c = \omega_0^2 m_e \epsilon_0/ e^2$,  $\omega_0$ the laser frequency, $\epsilon_0$ the vacuum permitivity, $m_e$ and $e$ the electron mass and charge). Such densities can be achieved with gas jets, that can be operated at high repetition rates ($>$ 1 kHz), are debris free, and produce high purity ion beams. The efficiency of the acceleration process(es) depends on different mechanisms associated with the laser-plasma interaction at near-critical density, including laser filamentation, electron heating, and beam filamentation driven by fast return currents. The density steepening on the front side of the target plays an important role in launching a shock capable of reflecting the slowly expanding background ions. Nevertheless, Fiuza \textit{et al.}\cite{Fiuza_2013} also demonstrated that the quality of the accelerated ion beam strongly depends on the density gradient on the rear side of the target. To obtain a narrow energy spread it is crucial to have a uniform shock velocity and ion reflection, which implies a uniform electron temperature profile, only achieved by a quick recirculation of the heated electrons due to the space-charge fields at the front and at the back of the target. Therefore, the target thickness should be limited to: $L_{target} < \lambda_0 (m_i/m_e)^{1/2}$. Considering a hydrogen plasma and a $\lambda_0$ = 1 $\mu$m laser implies $L_{target} < 40$ $\mu$m, and therefore relatively sharp gradients on both sides of the target. Other acceleration schemes at near-critical density \cite{Bulanov,Nakamura,Bulanov_comment,Wan} also require a sharp density gradient at the rear side of the (thin) target.
The acceleration process usually occurs in a short scale length, and once accelerated the dense ion-bunch may filament while propagating through the residual plasma/gas before reaching vacuum\cite{Mima,Davis,Gode}, which could degrade the final spatial quality of the beam as well as its shot-to-shot reproducibility. Suppressing/reducing this unnecessary part of the target could thus be very beneficial.

Ion acceleration by CSA, in a gas jet steepened at the entrance side of the driving beam, was first demonstrated \cite{Palmer_2011,Najmudin,Haberberger,Palmer_2015,Tresca,Chen-Y} using CO$_2$ lasers, the low critical density ($n_c \approx 10^{19}$ cm$^{-3}$) associated with their long wavelength ($\lambda_0$ = 10 $\mu$m) allowing to exploit regular-pressure, mm-scale gas jets. In these experiments, the laser pulse was a macro-pulse composed of several picosecond (ps) pulses. At the entrance of the gas jet the first pulses steepen the density gradient, so that the following pulses interact with a step-like density profile. This drives an electrostatic shock that reflects the ions, producing collimated proton beams with peaked energy spectrum and maximum energy of $\approx$ 20 MeV. For a better control of this density step, Tresca \textit{et al.}\cite{Tresca} used a single, low energy CO$_2$ ps-laser prepulse to drive a blast-wave (BW) inside the gas target, that generates the density gradient before the arrival of the collinear main ps-pulse\cite{Dover_2016}. For long density profiles ($\ge 40$ $\mu$m) broadband beams were produced, while for shorter plasma length ($\le 20$ $\mu$m) the energy spectrum was peaked and much narrower. Y. Chen \textit{et al.}\cite{Chen-Y} obtained similar results, but with a BW produced via a low energy nanosecond (ns) laser impacting a solid target placed at the entrance side of the gas jet.

Experiments using more widespread solid state ps-lasers ($\lambda_0 \approx 1 \mu$m, $n_c \approx 10^{21}$ cm$^{-3}$) have been performed using high-pressure gas jets \cite{Chen-SN,Puyuelo,Puyuelo-SPIE}, or exploded $\mu$m-size solid foils \cite{{Antici},{Pak}}. The plasma profile at the entrance side of the ps-pulse was pre-shaped by a nanosecond (ns) low-energy prepulse, allowing to explore the transition from TNSA to low density CSA. The number of accelerated protons was substantially higher ($\sim 10^4 \times$) than in experiments with CO$_2$ lasers conducted at lower density and smaller vector potential of the laser field.

Ion acceleration from a gas jet tailored both on its front and rear sides has been performed by Helle \textit{et al.}\cite{Helle}. Two counter-propagating hydrodynamic shock fronts were generated in a H$_2$ jet by laser ablating two opposite sides of the nozzle. At the intersection of these two BWs a $\sim$ 10-fold local density enhancement is generated with sharp ($\sim 30$ $\mu$m) gradients, allowing to reach near-critical density on a $\sim 70$ $\mu$m thickness. An ultraintense 800 nm, 50 fs laser pulse was then focused on this gas "wall". At $n_e = 0.6 n_c$ a wide-angle, low-energy beam typical of TNSA is observed, while at $n_e = 0.3 n_c$ a more focused beam with a high-energy halo is produced, attributed to Magnetic Vortex Acceleration (MVA) \cite{Bulanov}.

In order to tailor the gas jet, the hydrodynamic shock can be generated by laser-ablation of a solid target near the gas\cite{Kaganovich}, or by a ns heating-laser directly focused in the gas and collinear with the main ps laser beam\cite{Dover_2016,Passalidi,Puyuelo}. With the first method, the density is locally increased at the shock location, but the density profile behind and in front of the shock is weakly modified. With the second method, the heated region quickly expands, forming an evacuated cavity behind the shock front that significantly reduces the plasma density at the jet entrance. This offers the strong advantage of preserving the quality (intensity) of the main pulse before its arrival on the steep density gradient, increasing the efficiency of the acceleration process. In addition this method can be used at high repetition rate without damaging the nozzle and without debris production.  However, this method is difficult to realize experimentally for tailoring the density profile at the target rear side since it implies a counter propagating laser that could damage the laser chain, and makes more complicated the implementation of ion-beam diagnostics along the axis of the main laser, where the more energetic ions are expected.

To overcome this difficulties, we propose a new tailoring method based on a narrow gas jet coupled with two ns heating-lasers that propagate perpendicular to the main ps-laser pulse. This scheme can be implemented at high repetition rate, in a debris free environment. Using present high-density gas jets\cite{Henares}, it allows the creation of a thin ($\sim 10$ $\mu$m) plasma slab of adjustable density up to $2\times 10^{22}$ cm$^{-3}$.
The first part of this paper presents a study, based on hydrodynamic and on ion Fokker-Planck simulations, of this plasma tailoring method. In the second part, using kinetic (particle in cells) simulations, we explore the influence of the plasma shape on the production of high energy proton beams.

\section{Thin over-critical plasma production}\label{plasma tailoring}
\begin{figure}
	\includegraphics[width=\columnwidth]{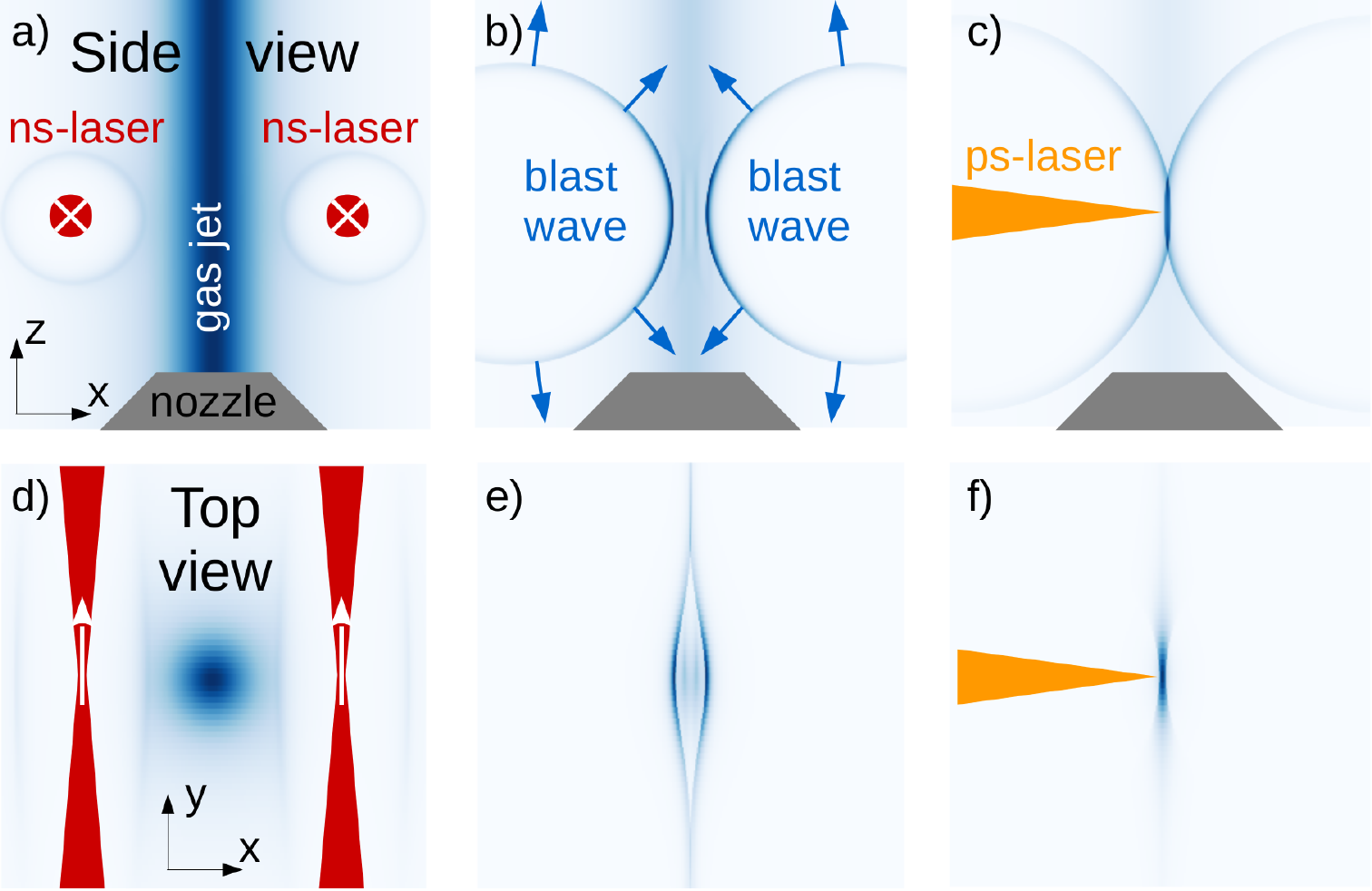}
	\caption{\label{Principe}\textit{Principle of the plasma tailoring. a)-c): side view (from y-axis) of the gas jet (blue column), that exits the nozzle along the z axis; d)-f) top view (from z-axis). The two synchronized, ns-beams (red) propagate parallel to the y axis, on both sides (x-axis) of the gas jet. The sudden laser-heating of the plasma induces steep density fronts (a,d) that propagate out of the laser axis as BWs (b,e). The collision of two BWs at the center of the gas jet (c,d) produces a sharp gradient, thin, high-density plasma.}}
\end{figure}

The principle of the tailoring method is illustrated in Figure \ref{Principe}. It is based on the use of a narrow, gaussian-like radial profile gas jet, coupled with two ns heating-lasers that propagate parallel to each other, and perpendicular to the main ps-laser. Both ns-beams are focused in the gas jet, one at the entrance side of the ps-beam, the other at the exit side. The ns-pulses are synchronized, and sent into the gas jet before the ps-pulse.

The sudden ionization and heating induced by each ns-beam generate a hydrodynamic shock that expels the plasma from the high intensity regions and produces a BW. In the post-shock region (behind the BW) the density is reduced, while a steep density profile is generated at the shock front, with a density increased by a factor $(\gamma +1)/(\gamma -1)$, where $\gamma$ is the heat capacity ratio of the gas. For a H$_2$ gas $\gamma = 1.4$, the density at the shock front is 6 times the background density. The absolute amplitude of this density peak is increasing for the front part that propagates toward the inside of the gas jet, while the rest disappears in the surrounding vacuum, leading to a step-like plasma profile. One can thus tailor at will the front, the rear, or both sides of the jet before the arrival of a driving ps-laser (for example for proton acceleration). The two counter-propagating BW collide at the jet center, leading to an increase of the central density by more than an order-magnitude, together with a reduction of the density in the vicinity of the sharp gradient(s). This scheme can be implemented at high repetition rate, in a debris free design. Using present high-density gas jets\cite{Henares}, it allows the creation of a thin ($\sim 10$ $\mu$m) plasma slab of adjustable density up to $2\times 10^{22}$ cm$^{-3}$. The typical lifetime of this transient target is $\sim$ 100 ps, hence the synchronization with the ps-pulse is easy to perform. 

To study the influence of the gas jet and laser parameters on the production of this thin transient plasma target, simulations were performed with the radiation-hydrodynamics code TROLL \cite{Lefebvre}, an arbitrary Lagrangian-Eulerian two- (2D) and three-dimensional (3D) radiation hydrodynamic code, with a ray-tracing package describing the laser propagation. 
Since many simulations are needed to explore our parameter space (laser energy and position, gas jet density and profile), and because supersonic gas jets have a gradient length along the jet axis ($z$ in Fig. \ref{Principe}) much longer than the radial gradient, we use a 2D geometry, in cartesian coordinates ($x-y$ plane). The simulation box is 1 mm $\times$ 1 mm. The gas column is at the center of the box ($x=0, y=0$) and corresponds to a supersonic hydrogen gas jet used in a previous experiment \cite{Puyuelo,Henares}, with a radial profile $n_{e0}(r) = n_0/(1+[r/r_0]^2)$, where $r_0$ = 70 $\mu$m. The gas is described using a perfect gas equation of state. Since hydrogen is very quickly ionized, we impose a full ionization at the beginning of the simulation. However, in order to prevent an early gas jet expansion, we impose an initial temperature of 300 K. Each heating-beam propagates along the $y$-axis at a distance $\pm \Delta x$ from the jet center, and is focused at $y=0$ in a Gaussian focal spot of 15 $\mu$m (full width at half maximum, FWHM), has a wavelength of 1 $\mu$m (corresponding to a critical density $n_c = 10^{21}$ cm$^{-3}$), a Gaussian temporal profile of FWHM = 0.4 ns, with its maximum at $t=0.7$ ns.

\subsection{Tailoring from a single beam}\label{Single_beam_tailoring}
\subsubsection{General behavior}
\begin{figure}
	\includegraphics[width=\columnwidth]{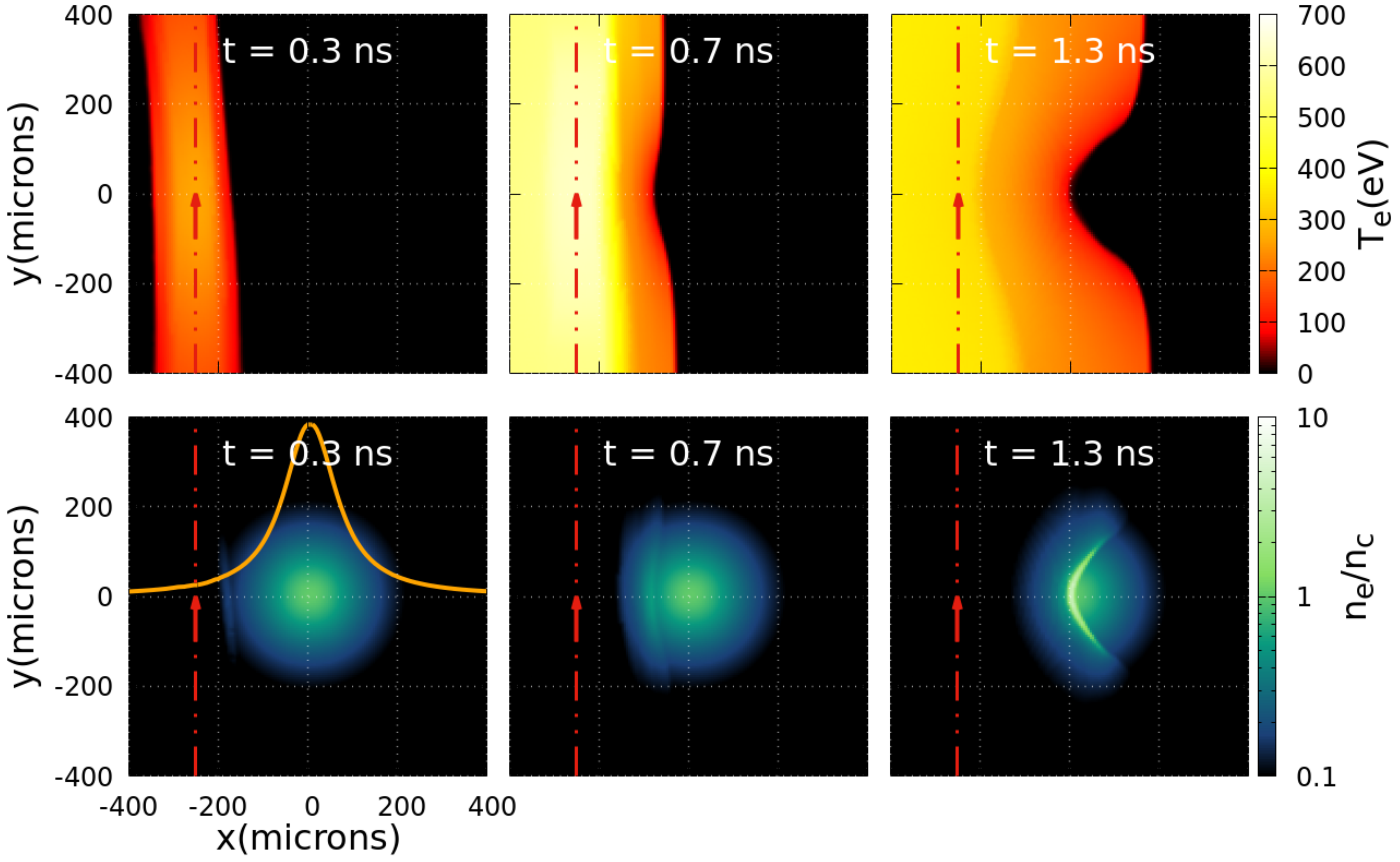}
	\caption{\label{1faisceau_2D}\textit{2D-profiles of $T_e$ (top graphs) and $n_e$ (bottom graphs) at 3 different times (left to right), for a H$_2$ gas jet tailored by a single ns-beam propagating along $y$-axis at $x$ = -250 $\mu$m and focused at $y$ = 0 $\mu$m (red arrow). The initial peak density is $n_0$ = $n_c$, the orange curve in the bottom-left graph is the normalized profile at $y=0$. The incident laser energy is 50 J ($I_{max} = 4.6\times 10^{16}$ W/cm$^2$).}}
\end{figure}

An example of plasma tailoring with only one beam is presented in figure \ref{1faisceau_2D}. The laser and plasma parameters are detailed in the figure caption. The 2D spatial profiles of the electron temperature $T_e$ (top graphs) and density $n_e$ (bottom graphs) are shown at 3 times. The Rayleigh length of the laser is relatively long, $z_R \sim 250$ $\mu$m, leading to a homogeneous heating of the plasma along the propagation axis ($y$). The thermal pressure associated with this sudden heating rapidly expels transversely ($x$-axis) the plasma from the high-intensity regions, driving a hydrodynamic shock. The bottom row shows that the shock front starts near the laser axis, and propagates toward the gas jet center, where the amplitude of the density perturbation reaches its maximum, $n_e^{front} \sim 5 n_c$, close to the factor 6 expected for an adiabatic compression. Despite the homogeneous energy deposition along the laser axis, the shock propagates faster at the edge of the gas jet than in its center ($y=0$), bending its front during its propagation, and leading to the bow shape observed at $t$ = 1.3 ns. This velocity difference is induced by the density profile of the gas jet: the velocity of the shockwave is higher in the lower density regions of the gas jet\cite{Kaganovich}. Notice that in the temperature profile at $t$ = 0.3 ns one can also see that despite the much lower density at the edge of the gas jet ($n_e(r=\Delta x) \sim 0.07 n_c$), the laser beam is slightly deflected by the density gradient. In the context of proton acceleration by a driving ps-pulse, this illustrates the importance of reducing as much as possible the plasma density in front of the sharp density gradient.

The energy absorbed in the plasma is $\sim$ 240 mJ, which is only 0.47 $\%$ of the incident one. It corresponds to the heating of the plasma column by inverse bremsstrahlung absorption: centered on the laser axis, this column has a density $n_e \sim 7\times 10^{19}$ cm$^{-3}$, a length of 1 mm (size of the simulation box along $y$), a radius of $\sim 100$ $\mu$m (Fig. \ref{1faisceau_2D}), and is heated to $\sim$ 700 eV. Such a heated volume has an energy of $\sim$ 250 mJ, in good agreement with the absorbed energy.

\begin{figure}
	\includegraphics[width=\columnwidth]{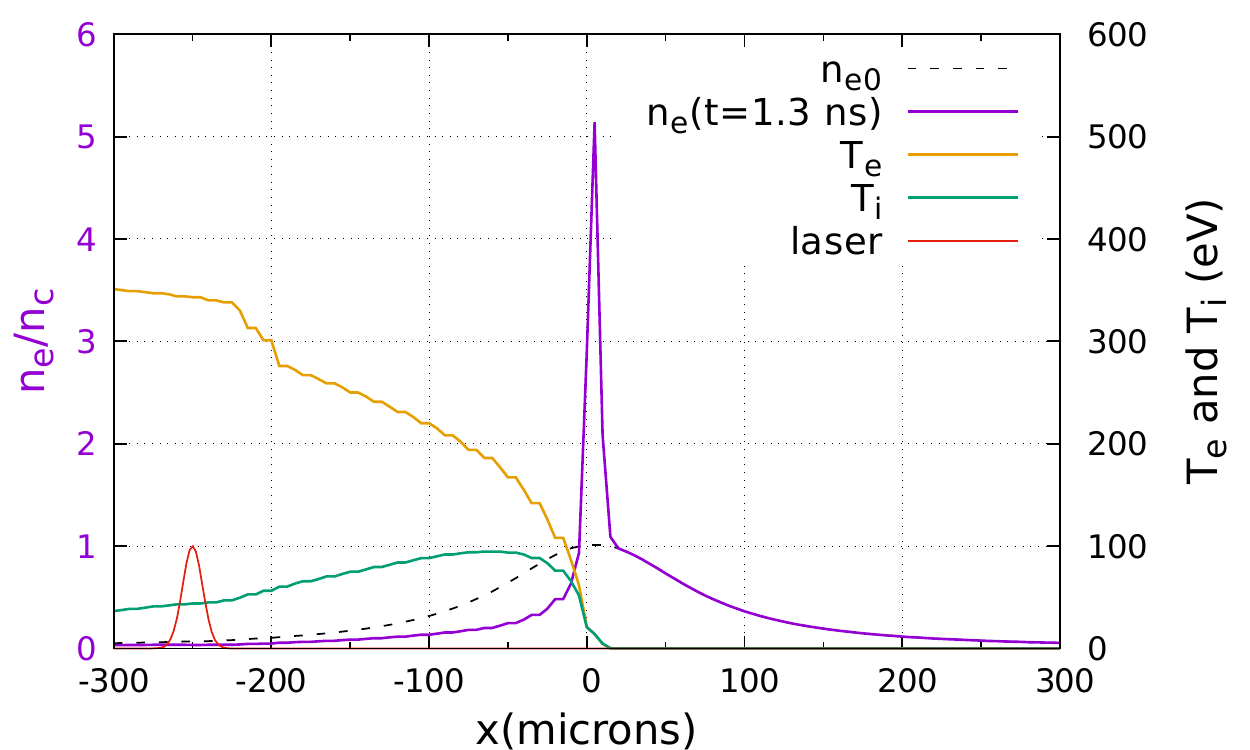}
	\caption{\label{1faisceau_coupes}\textit{Profiles of density ($n_e/n_c$, violet curve) and temperature ($T_e$: orange curve; $T_i$: green curve) obtained from Fig. \ref{1faisceau_2D} at $y=0$ and $t=1.3$ ns. The red curve is the normalized profile of the laser focal spot. The black-dashed line is the initial density profile.}}
\end{figure}

Line-outs of the electron and ion temperatures ($T_e$ and $T_i$ respectively) and density profiles at $t$=1.3 ns (when the shock front reaches the jet center) are shown in figure \ref{1faisceau_coupes}. Notice that not only the initial peak density is increased by a factor $\sim 5$, but the pedestal of the density profile on the sharp gradient side ($x < 0$) is also significantly reduced. In addition, the plasma temperature in this region is relatively high (100s eV), which reduces the inverse bremsstrahlung coefficient and thus the absorption that could undergo a driving ps-pulse for proton acceleration. This perpendicular tailoring technique can produce density profiles quite similar to those obtained from the scheme based on a low-energy prepulse collinear to the main driving beam\cite{Dover_2016,Passalidi,Puyuelo}, it completely decouples the tailoring ns-beam from the driving ps-beam, and offers another parameter of adjustment, the "impact parameter" $\Delta x$.

\begin{figure}
	\includegraphics[width=\columnwidth]{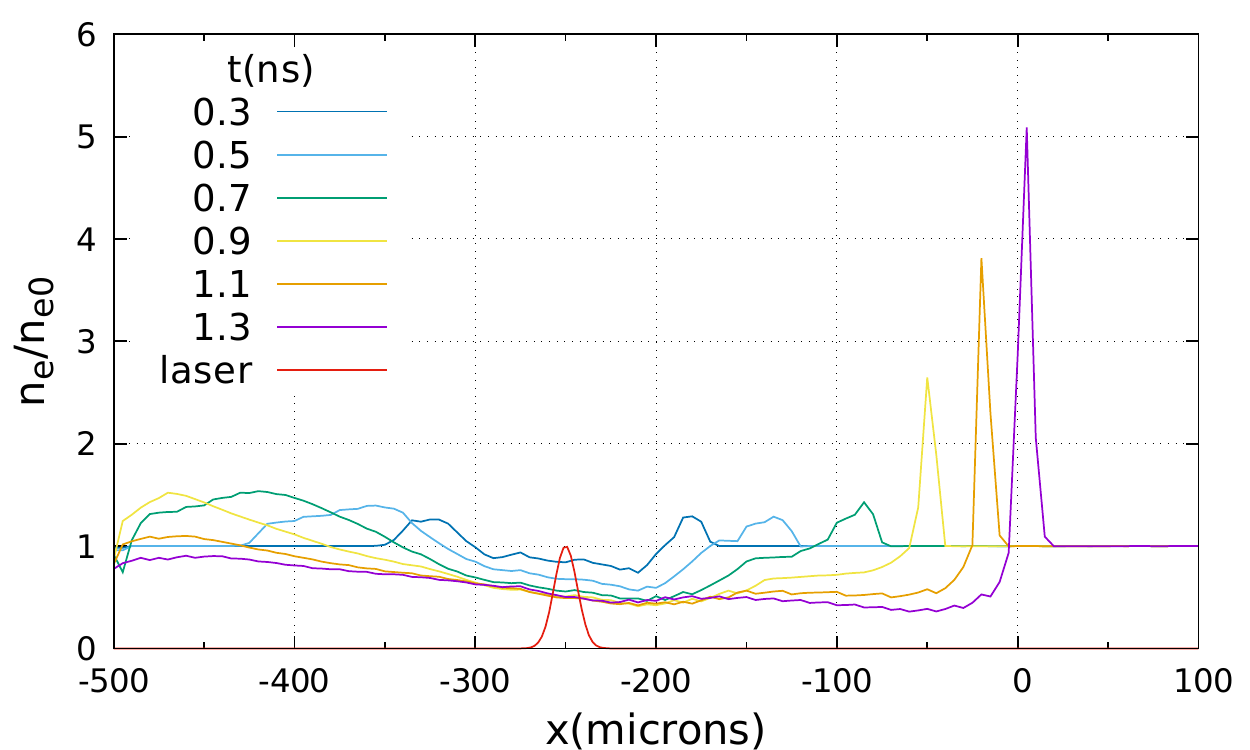}
	\caption{\label{1faisceau_coupes_fcn_temps}\textit{Evolution of the relative density perturbation $n_e/n_0$, at $y=0$. The red curve is the normalized profile of the laser focal spot. From same simulation than Fig. \ref{1faisceau_2D}.}}
\end{figure}

The evolution of the relative density perturbation $n_e/n_{e0}$ is presented in figure \ref{1faisceau_coupes_fcn_temps}. Due to the velocity dispersion of the shock front in the background density profile, the front moving downward the density gradient (left in Fig. \ref{1faisceau_coupes_fcn_temps}) accelerates and becomes broader and weaker, while the front moving upward slows down and becomes continuously sharper and stronger, until the maximum compression value is reached, $n_e/n_{e0} = (\gamma+1)/(\gamma-1)$.

\begin{figure}
	\includegraphics[width=\columnwidth]{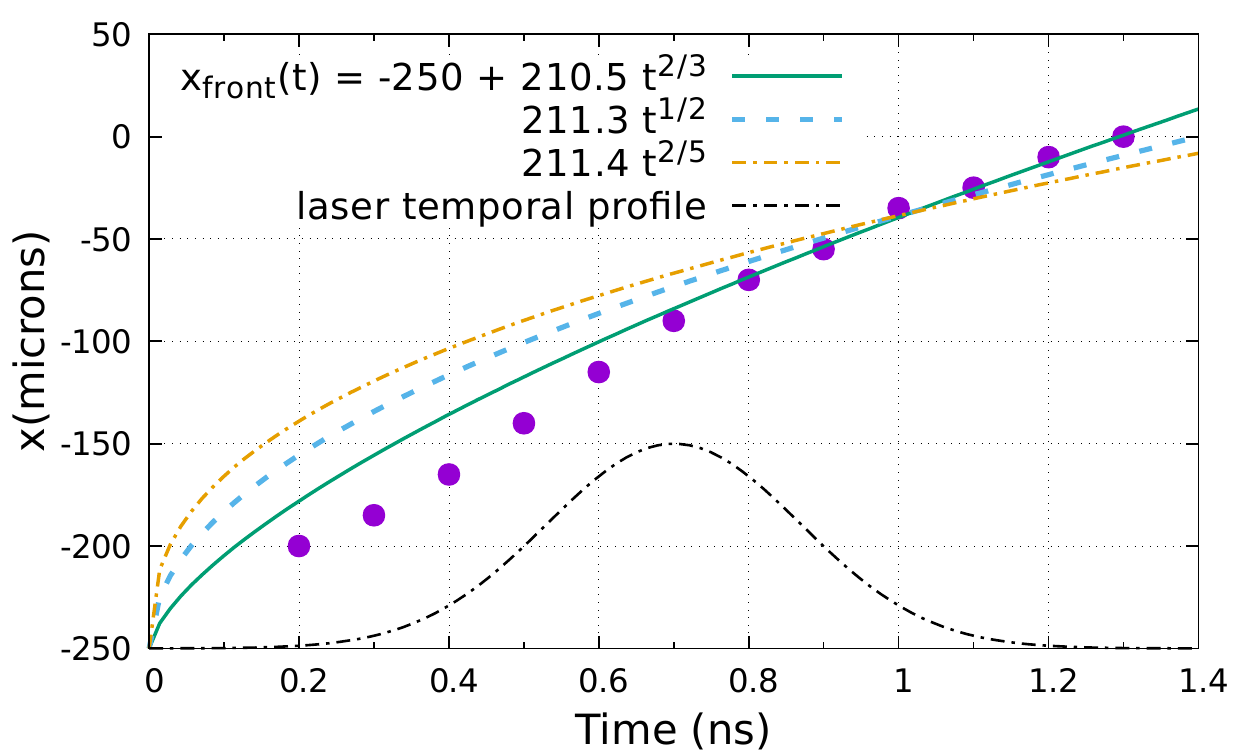}
	\caption{\label{Distance_BW_fcn_temps}\textit{Position of the shock-front versus time, at $y=0$. The curves are fits of the front position for $t\ge 0.7$ ns (after the laser pulse maximum) using the expansion law given by Eq. \ref{r_BW} with $\alpha$ = 1 (green curve), 2 (blue-dash curve) and 3 (orange-dash-dot). The black dash-dot curve is the normalized temporal profile of the laser. From same simulation than Fig. \ref{1faisceau_2D}.}}
\end{figure}

The position, as a function of time, of the shock front that propagates toward the jet center is presented in figure \ref{Distance_BW_fcn_temps}. After the laser pulse maximum, when most of the energy transfer from the laser to the plasma has occurred (see following), the shock front is clearly formed and has left the heated region (see figures \ref{1faisceau_2D} and \ref{1faisceau_coupes_fcn_temps}), the evolution of the shock front is similar to the well known Sedov-Taylor solution for an adiabatic blast wave\cite{Sedov,Taylor,Sedov_book,ZelDovich}. This solution describes a shock-wave in which energy, momentum and mass behind the shock front are conserved, causing the initial energy of the wave to be spread over an increasingly larger volume, resulting in a blast wave radius that varies with time as:

\begin{equation}\label{r_BW}
r_{BW}(t) = \zeta(\gamma,\alpha) \left(\frac{E_0}{\rho}\right)^{\frac{1}{2+\alpha}} t^{\frac{2}{2+\alpha}}
\end{equation}

\noindent where $\alpha$ is the dimensionality of the shock ($\alpha$=1, 2, 3 for, respectively, a planar, cylindrical, or spherical shock), $E_0$ is the energy released, $\rho$ the mass density of the ambient medium ($\rho = A m_p n_e / Z$; $Z$, $A$, the ion charge state, and mass number, and $m_p$ the proton mass, respectively), and $\zeta(\gamma,\alpha)$ a constant close to one ($\zeta$=0.98 for H$_2$ and cylindrical expansion). The curves in Fig. \ref{Distance_BW_fcn_temps} are fits of the simulations points for $t\ge 0.7$ ns (after the laser pulse maximum), using the expansion law given by Eq. \ref{r_BW}, with $\alpha$ = 1 (green line), 2 (blue-dashed line) and 3 (orange-dotted line). The Rayleigh length of the ns-beam is $z_R \sim 250$  $\mu$m, so that the high-intensity region covers the entire length of the gas jet (along the $y$-axis in Fig. \ref{1faisceau_2D}). In a real 3D-geometry, since $z_R$ is larger than the gas jet radius, the expansion would be cylindrical ($r_{front} \propto t^{1/2}$), centered on the laser axis. However, due to the 2D planar geometry of the present simulations, the shock front follows a planar expansion law, $x_{front} \propto t^{2/3}$ (green line).

\subsubsection{Dependence on the incident laser energy}
\begin{figure}
	\includegraphics[width=\columnwidth]{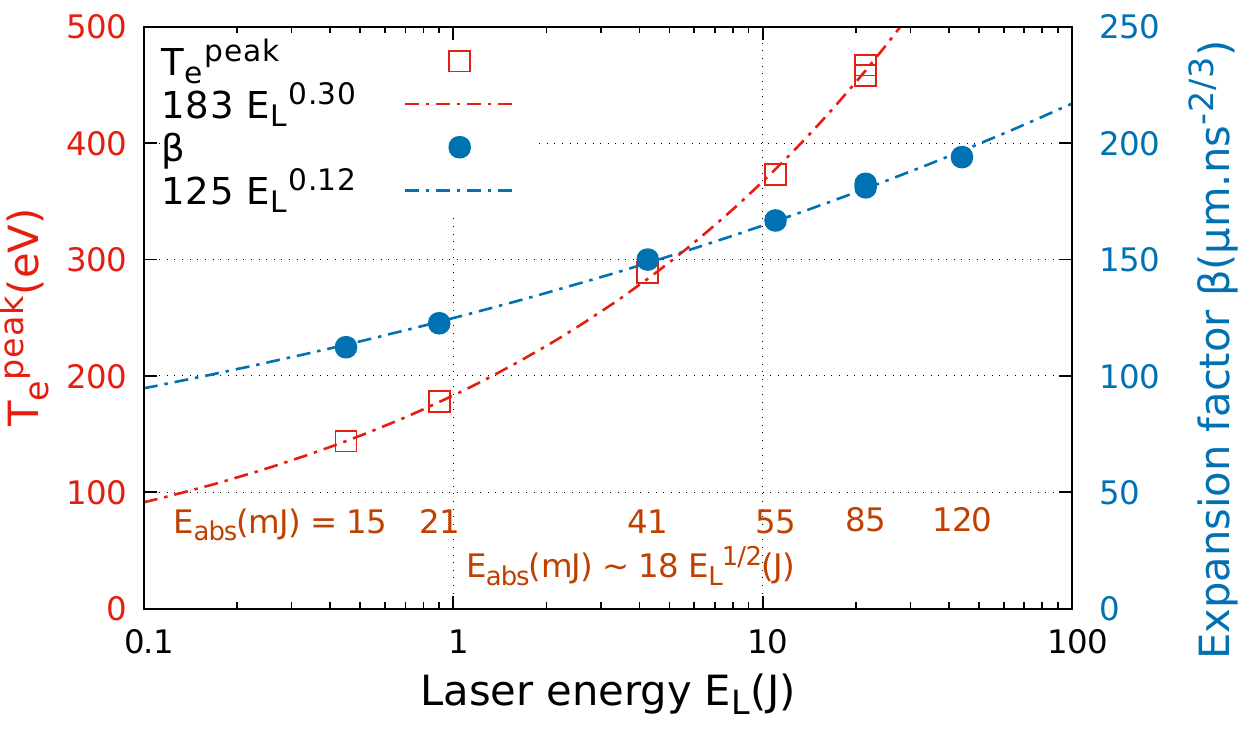}
	\caption{\label{Eabs_expansion_vs_Elaser}\textit{Evolution with the incident laser energy $E_L$. Red, left scale: peak value of $T_e$ (reached on the laser axis, at the pulse maximum). Blue, right scale: factor $\beta$ of the radial expansion $r(t) = \beta t^{2/3}$. The dashed curves are fits. Values in dark-orange: laser energy absorbed $E_{abs}$ by the plasma, for each set of ($T_e$,$\beta$) points. A fit of these values gives $E_{abs}(mJ) \sim 18$ $E_L^{1/2}(J)$. As in Fig. \ref{1faisceau_2D}, the initial peak density is $n_0/n_c$ = 1, and the laser propagates along $\Delta x$ = 250 $\mu$m.}}
\end{figure}

The amplitude and the velocity of the BW depend on energy released $E_0$ (Eq. \ref{r_BW}), which in turn depends on the plasma heating by the laser, and thus on the incident laser energy $E_L$. This one needs to be high enough to quickly heat the plasma and generate a strong hydrodynamic shock. In the laser-plasma conditions of Fig. \ref{1faisceau_2D} ($n_0/n_c$ = 1, $\Delta x$ = 250 $\mu$m), the maximum compression factor at the shock front, $n_e^{front}/n_e$, is $\sim$ 3 at $E_L \sim$ 0.5 J, reaches $\sim$ 4 at $E_L \sim$ 5 J, and "stabilizes" to a peak value of 5 for $E_L >$ 10 J, which is close to the maximum value $(\gamma +1)/(\gamma-1) \sim 6$ for adiabatic compression of H$_2$. In figure \ref{Eabs_expansion_vs_Elaser} the dependence on the laser energy, of $T_e$ and of the factor $\beta$ that governs the BW expansion, $r(t) = \beta t^{2/3}$ (Eq. \ref{r_BW}), are shown. As expected, the larger $E_L$, the is larger the energy deposition $E_{abs}$, and thus $T_e$. More energy is transferred to the BW, that expands faster. This transfer is however not linear: in the present case, $T_e$ grows like $E_L^{0.30}$, and $\beta$ increases like $E_L^{0.12}$. A change of the laser energy by an order of magnitude only changes the expansion rate by a factor 1.3. Amplitude and velocity of the BW are thus weakly dependent on the laser energy, making this tailoring scheme very robust to the laser fluctuations.

\subsubsection{Dependence on the initial plasma density}
\begin{figure}
	\includegraphics[width=\columnwidth]{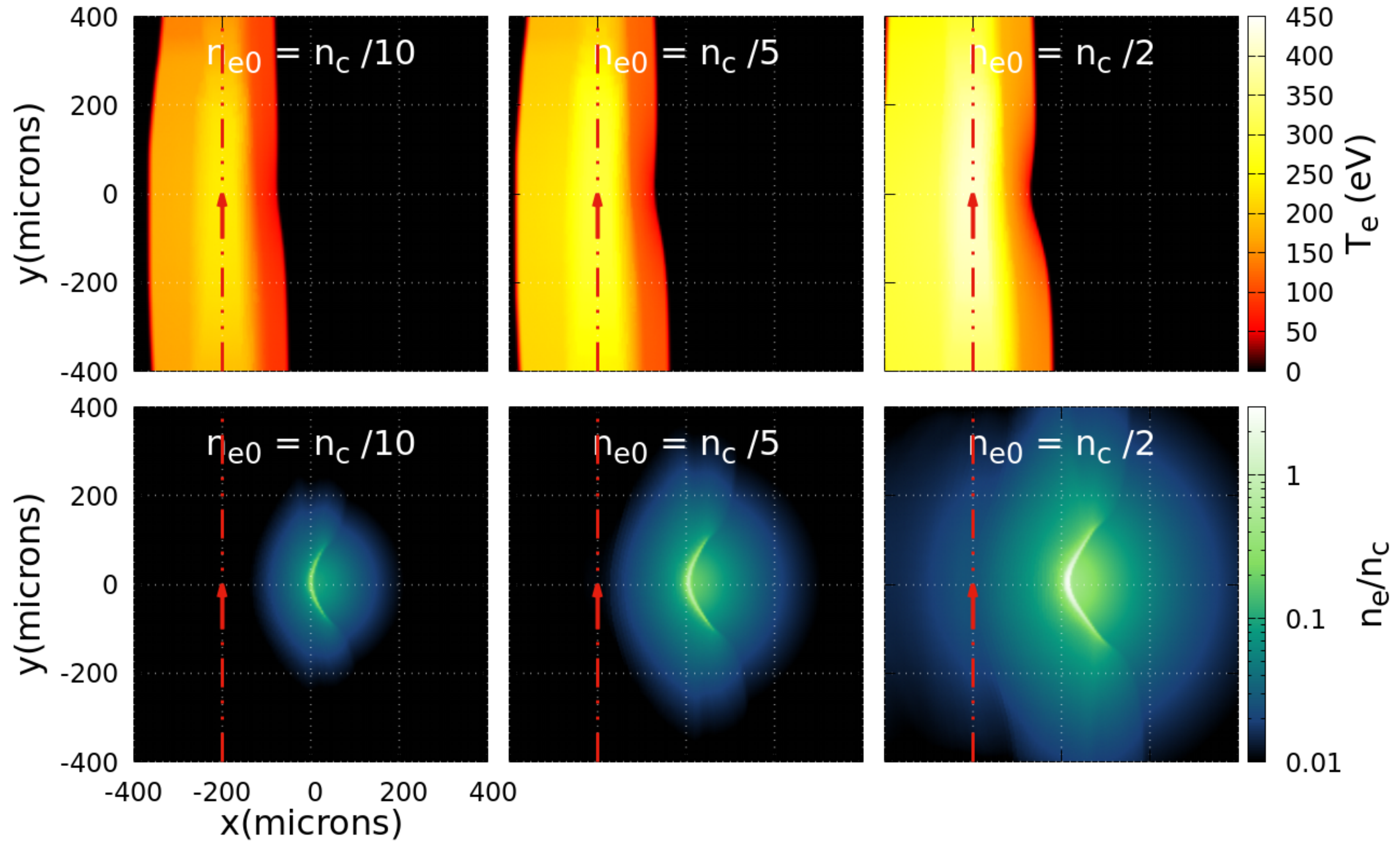}
	\caption{\label{2D_profiles_1beam_vs_ne0}\textit{2D-profiles of $T_e$ (top graphs) and $n_e$ (bottom graphs) at 3 different densities ($n_0$), for a H$_2$ gas jet. The ns-beam propagates at $\Delta x$ = $-200 \mu$m (red arrow) and has an energy of 25 J ($I_{max} = 2.3\times 10^{16}$ W/cm$^2$). $T_e$ profiles are taken at $t$ = 0.7 ns (laser pulse maximum), $n_e$ profiles at $t$ = 1.4 ns (shock front at the jet center).}}
\end{figure}

\begin{figure}
	\includegraphics[width=\columnwidth]{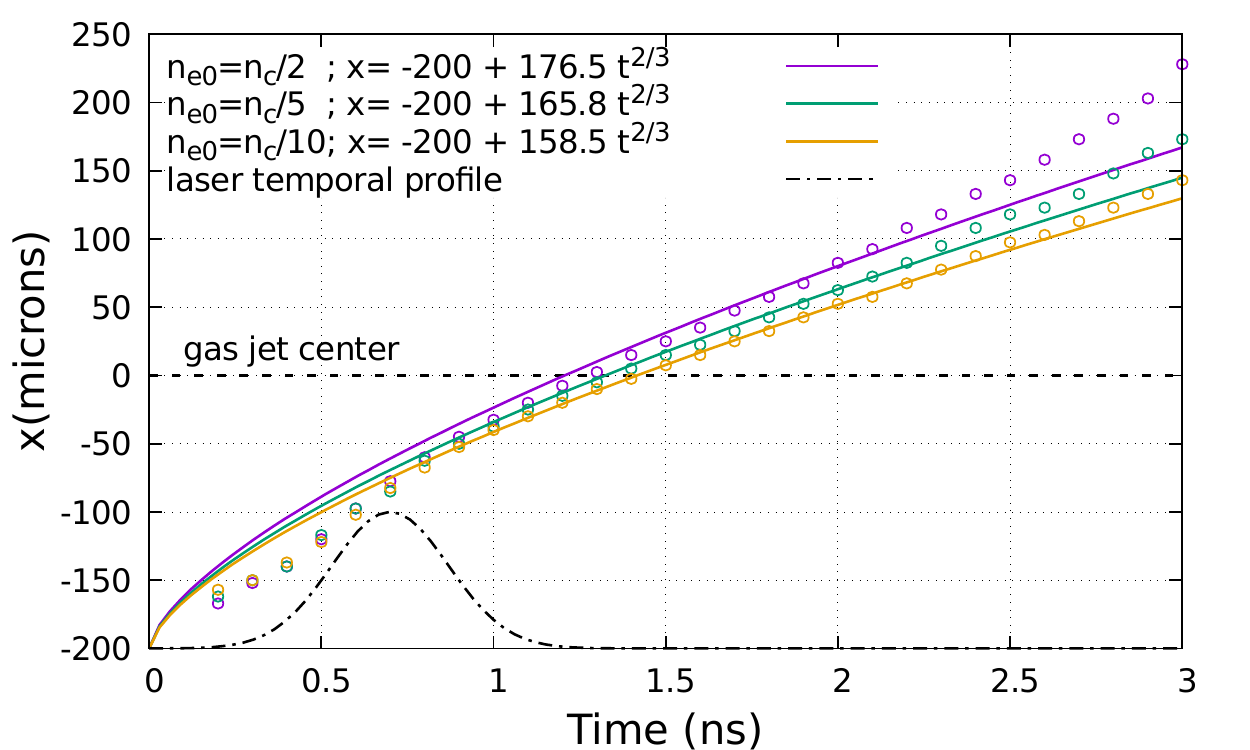}
	\caption{\label{Evolution_position_vs_density}\textit{Position of the shock-front versus time, at $y=0$, and for three initial density $n_0$, from the same simulations than Fig. \ref{2D_profiles_1beam_vs_ne0}. The curves are fits using the expansion law given by Eq. \ref{r_BW} with $\alpha$ = 1. The black dash-dot curve is the laser temporal profile.}}
\end{figure}

Figure \ref{2D_profiles_1beam_vs_ne0} presents, for three different initial plasma densities ($n_0/n_c$ = 1/10, 1/5, and 1/2),  the spatial profiles of the electron temperature at the laser pulse maximum ($t=0.7$ ns, top row), and of the electron density when the shock arrives at the jet center ($t=1.4$ ns, bottom row). The initial laser conditions are slightly different from simulation in Fig. \ref{1faisceau_2D}: the laser beam is set closer to the gas jet ($\Delta x$ = -200 $\mu$m) and has a two times lower energy (25 J). The top graphs show that the larger the initial plasma density $n_0$, the hotter is the plasma. At the maximum of the laser ($t=0.7$ ns), $T_e(y=0) \sim 479 (n_0/n_c)^{1/3}$, and at the laser exit ($y=1$ mm) the total energy transferred to the plasma is $E_{abs}$(mJ)$\sim 120 (n_0/n_c)^{3/2}$. Nevertheless, the bottom graphs show that despite an expected larger energy $E_0$ transferred to the BW, the velocity of the shock front is weakly modified: the gas jet center is reached at almost the same time ($t \sim 1.4$ ns) for these three initial densities. This is the consequence of the term $\rho^{-1}$ in Eq. \ref{r_BW}, that counter-balances part of the increase of $E_0$. Assuming $E_0 \propto E_{abs}$, the shock front should globally evolve like $n_0^{1/6}$ in the planar geometry of these simulations, and like $n_0^{1/8}$ in a real 3D configuration. Figure \ref{Evolution_position_vs_density} shows the evolution of the longitudinal position $x$ of the shock front, at $y=0$, for the three initial densities $n_0$ of Fig. \ref{2D_profiles_1beam_vs_ne0}. Similarly to Fig. \ref{Distance_BW_fcn_temps}, the propagation of the front in the core of the gas jet follows the expansion law of a BW given by Eq. \ref{r_BW} (with $\alpha$ = 1). It also illustrates a weak dependence on $n_0$: the expansion coefficient of the fits follows $183.4 (n_0/n_c)^{0.063}$ $\mu$m.ns$^{-2/3}$, a power law even weaker than expected (assuming $E_0 \propto E_{abs}$).

\begin{figure}
	\includegraphics[width=\columnwidth]{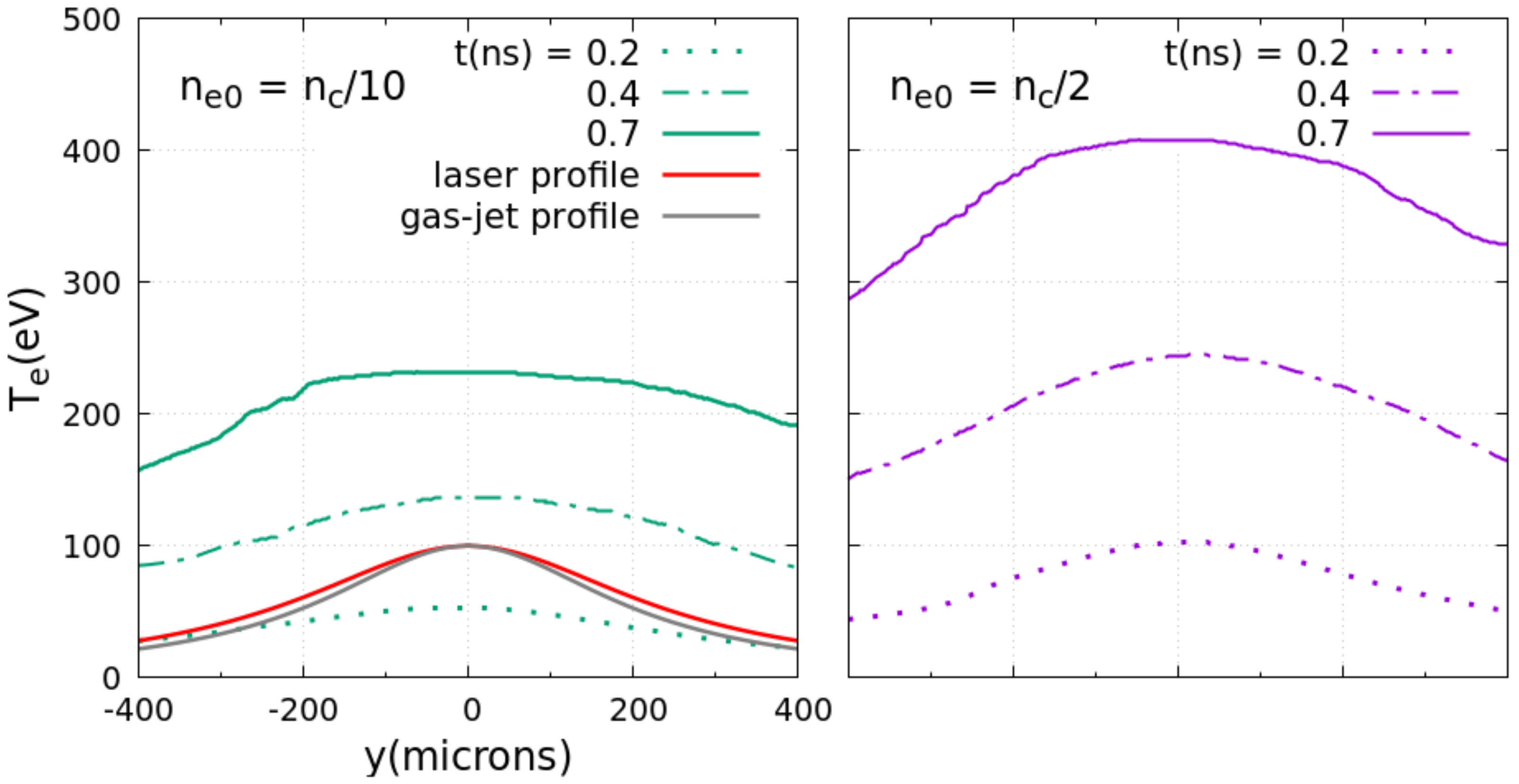}
	\caption{\label{Te_1beam_vs_ne0_time}\textit{Line-outs of $T_e$ along the laser axis $y$ ($x$ = -200 $\mu$m) of the simulations in Fig. \ref{2D_profiles_1beam_vs_ne0}, for $n_0/n_c$ = 1/10 and 1/2, and at $t=$ 0.2, 0.4, and 0.7 ns (laser maximum). The red and gray curves are respectively the normalized profiles of the laser and the gas jet.}}
\end{figure}

As previously shown, the velocity of the shock front at $y=0$ weakly depends on $n_0$ (the background gas jet density). However, the bottom graphs in Fig. \ref{2D_profiles_1beam_vs_ne0} show, as Fig. \ref{1faisceau_2D}, a bow shape of the front when it reaches the center of the gas jet. This difference in behavior comes from the homogeneous energy deposition along the laser axis, so that $E_0$ is $\sim$ constant along the $y$-axis ($x=-200$), and it is only the density profile ($\rho$ in Eq. \ref{r_BW}) that drives the velocity of the shock front, leading to a faster expansion on the edge of the gas jet, where the density is lower and has a smoother gradient.
Figure \ref{Te_1beam_vs_ne0_time} shows lineouts of $T_e$ along the laser axis, at two densities ($n_0/n_c$ = 1/10 and 1/2), and at $t=$ 0.2, 0.4, and 0.7 ns (laser maximum). While the laser intensity $I_L$ and the plasma density are $\sim$ 2 times lower at $y=\pm 250$ $\mu$m, the temperature profile becomes rapidly homogeneous along the laser axis: the bremsstrahlung absorption coefficient is proportional to $T_e^{-3/2}$, leading to a "saturation" of the heating in the high intensity regions. The simulation results presented in Fig. \ref{Eabs_expansion_vs_Elaser} show that $T_e^{peak}$ evolves like $I_L^{0.3}$, in good agreement with profiles shown in Fig. \ref{Te_1beam_vs_ne0_time}.

To summarize, the choice of the initial density $n_0$ (backing pressure of the gas jet), has almost no effect on the compression factor at the shock front, $n_e ^{front}/n_{e0} \sim (\gamma+1)/(\gamma-1)$, nor its spatial profile (bow shape), and only increases linearly the absolute peak density. These conclusions are only valid in the absence of strong beam refraction, which could occur if the ns-beam propagates in a high density region, at small $\Delta x$ or too high $n_0$.

Dependence of the plasma tailoring on the initial position $\Delta x$ of the laser is presented later in section \ref{dependence_on_distance}.

\subsubsection{Plasma profile at late times}
\begin{figure}
	\includegraphics[width=\columnwidth]{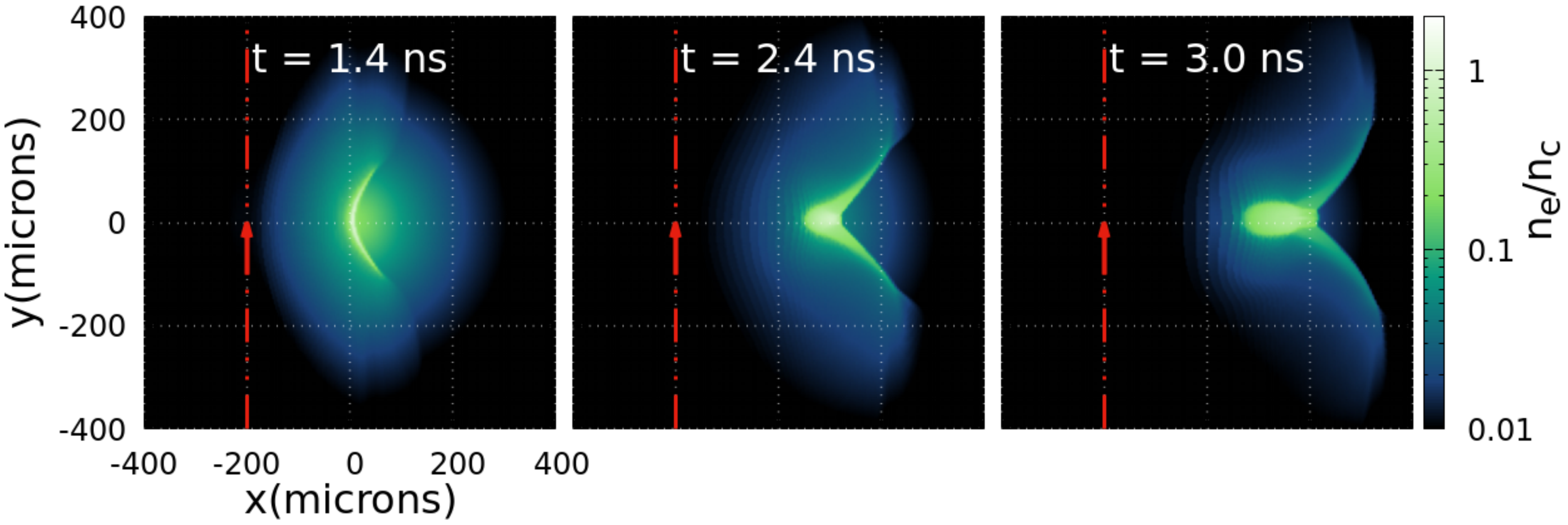}
	\caption{\label{2D_1beam_late_time}\textit{2D-profiles of $n_e$ at three consequent times (left to right). From the simulation in Fig. \ref{2D_profiles_1beam_vs_ne0} at $n_0 = n_c/5$.}}
\end{figure}

\begin{figure}
	\includegraphics[width=\columnwidth]{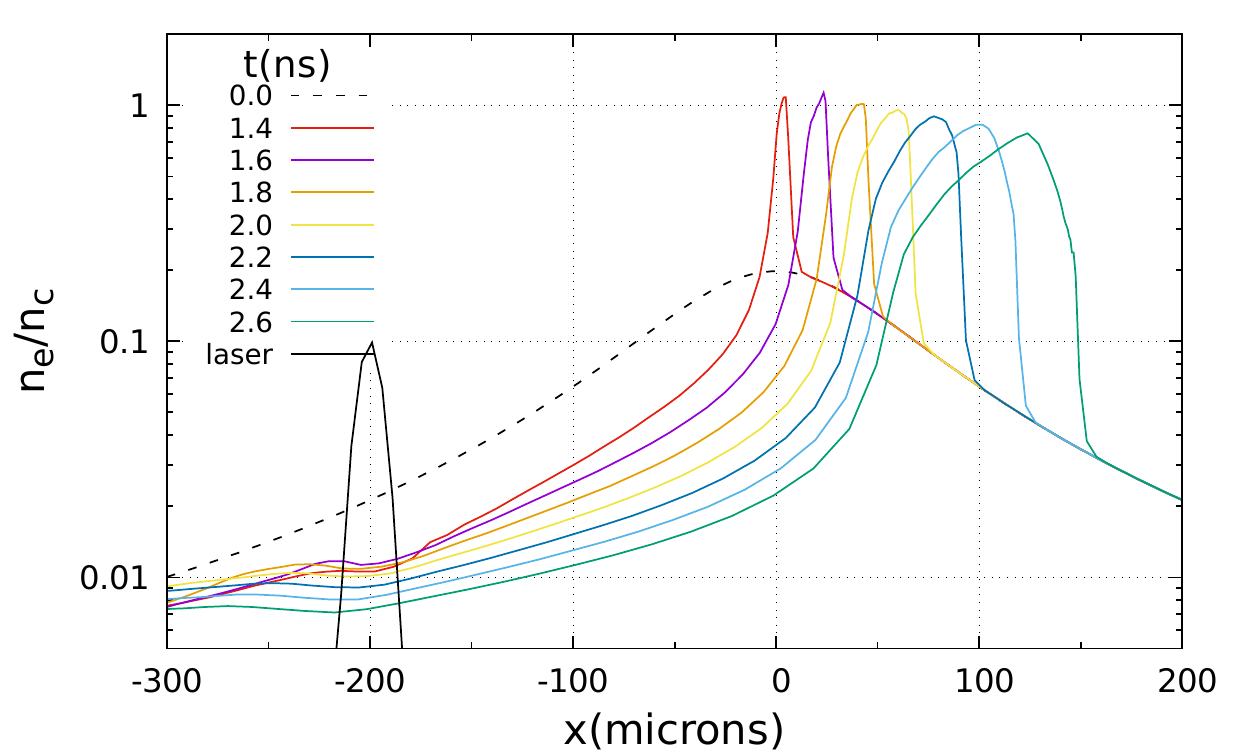}
	\caption{\label{Coupes_1beam_late_time}\textit{Line-outs of the plasma electron density profiles obtained from the same simulation as in Fig. \ref{2D_1beam_late_time} at $y=0$. The black curve is the normalized profile of the laser focal spot. The black-dashed line is the initial density profile. From the simulation in Fig. \ref{2D_1beam_late_time}}}
\end{figure}

Figure \ref{2D_1beam_late_time} shows the evolution of the density profile when the shock front overcomes the gas jet center. Once the shock front reaches the down ramp, it broadens along $x$, while the peak density remains almost at the same level, as can be observed on the line-outs in Fig. \ref{Coupes_1beam_late_time}. The plasma thickness increases by almost a factor 10 while the peak density only decreases from $n_c$ to 0.65$n_c$, and still with relatively sharp gradients. This is the result of the bow shape evolution that tends to transversally ($y$-axis) compress the plasma from the gas jet edge toward the propagation axis ($x$-axis). This happens thanks to the (cylindrical) shock that is excited all along the laser axis, while the gas jet radius is smaller than the laser Rayleigh length. In the case of the previously studied schemes based on an axial plasma shaping \cite{Dover_2012,Passalidi,Puyuelo}, the shock starts from (or close to) the gas jet center, and the spherical expansion does not offer a transverse compression of the shock. With perpendicular plasma tailoring, the induced transverse compression offers a way to produce near- or over-critical plasmas of adjustable thickness. 

\subsection{Tailoring both sides of the gas jet}

\subsubsection{General behavior}
In the quest of producing the thin and dense plasma slabs with very steep gradients required for numerous applications such as proton acceleration, tailoring the plasma from two opposite sides appears as a promising solution. To preserve the main driving beam from detrimental effects such as filamentation, refraction, absorption, it is essential to reduce as much as possible the plasma wings in front of the density gradient. However, the compression factor associated to a single BW cannot be larger than $(\gamma+1)/(\gamma-1)$, corresponding to 6 for H$_2$ and 4 for He. A possible way to increase this compression factor is to collide two blast waves. For a given required peak density this allows reducing the initial density $n_0$, and thus the wing level, by at least a factor two, and possibly more if the collision occurs in a regime of stagnation.

\begin{figure}
	\includegraphics[width=\columnwidth]{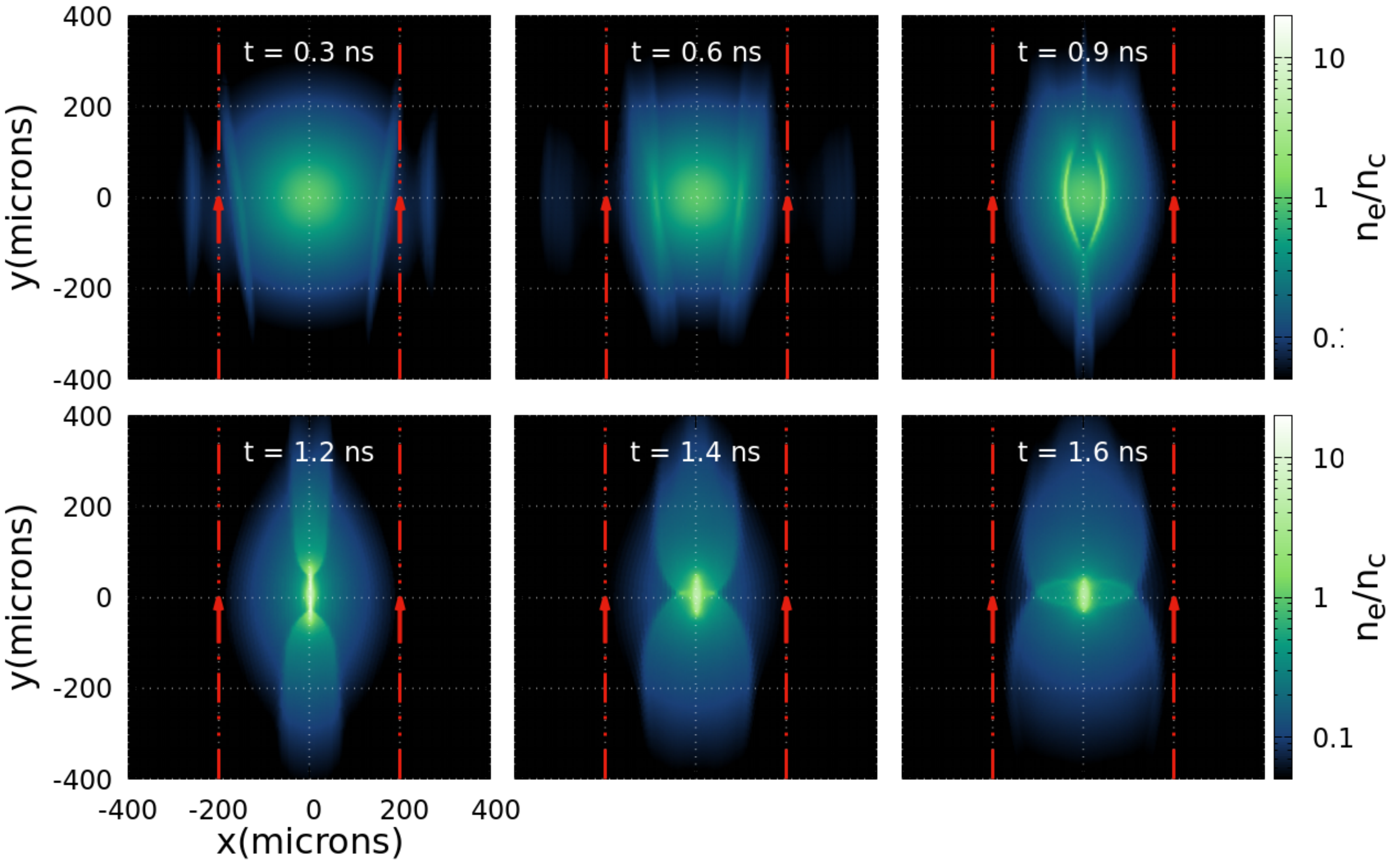}
	\caption{\label{Profil_2D_2faisceaux_fcn_temps}\textit{2D density profiles at different times, for a plasma tailored by two parallel ns-beams. The initial density is $n_0/n_c$ = 1, and the laser parameters are the same as in Fig. \ref{2D_profiles_1beam_vs_ne0} and \ref{2D_1beam_late_time}. The two beams propagate at $\Delta x$ = $\pm 200$ $\mu$m from the jet center (red arrows).}}
\end{figure}

An example of plasma tailoring by two parallel ns-beams is presented in figure \ref{Profil_2D_2faisceaux_fcn_temps}. The initial density is $n_0/n_c$ = 1, the laser parameters of each beam are the same as in Fig. \ref{2D_profiles_1beam_vs_ne0} and \ref{2D_1beam_late_time}, and the beams propagate on both sides of the gas jet, at $\Delta x = \pm$ 200 $\mu$m from the jet center. Two counterpropagting BW are generated simultaneously, each with an amplitude quickly reaching $\sim$ 5-6 times the local density. 

At the beginning of the BWs generation, $t$=0.3 ns in Fig. \ref{Profil_2D_2faisceaux_fcn_temps}, due to the high initial density ($n_e\sim 0.07 n_c$ at the laser axis), the ns-beams are deflected on the density gradient of the edge of the gas jet, leading to BW fronts initially tilted. However, the angle between the two fronts reduces along their propagation. Due to the bow shape of BWs, their collision at $x=0$ starts at $t$ = 0.8-0.9 ns from the edges of the BWs' fronts, $y\sim - 200 \mu$m. The central parts ($y=0$) of the shock fronts collide at $t$ = 1.2 ns, producing a plasma of near-rectangular shape of $\sim$ 10-20 $\mu$m thickness ($x$-axis) and $\sim$ 100 $\mu$m width ($y$-axis). This transient plasma has very sharp gradients on both sides, and a high peak density of $\sim$ 15 $n_c$. Such a value is impossible to reach by tailoring plasma by a single laser beam. 

At later times (after the jet center is reached), in the case of a single BW, as discussed previously (Fig. \ref{2D_1beam_late_time} and \ref{Coupes_1beam_late_time}), the central high-density plasma becomes rapidly thicker and continues its motion toward the other side of the gas jet. In the case of two counter-propagating BWs, Fig. \ref{Profil_2D_2faisceaux_fcn_temps} shows that, the (fast traveling) edges of the BWs ($|y| >$ 100 $\mu$m) quickly interpenetrate and continue their opposite motions toward the other side of the jet, while the collision near $x=0$ lasts much longer, resulting in a high-density region that stagnates. In the laser-plasma conditions considered in these simulations, the velocity of the BW along the $x$-axis is $\sim$ 100 $\mu$m/ns, so that the typical duration of the collision, the lifetime of the sharp-gradient transient plasma, is a few hundred of picoseconds.

\subsubsection{Influence of the distance between the two lasers}\label{dependence_on_distance}
\begin{figure}
	\includegraphics[width=\columnwidth]{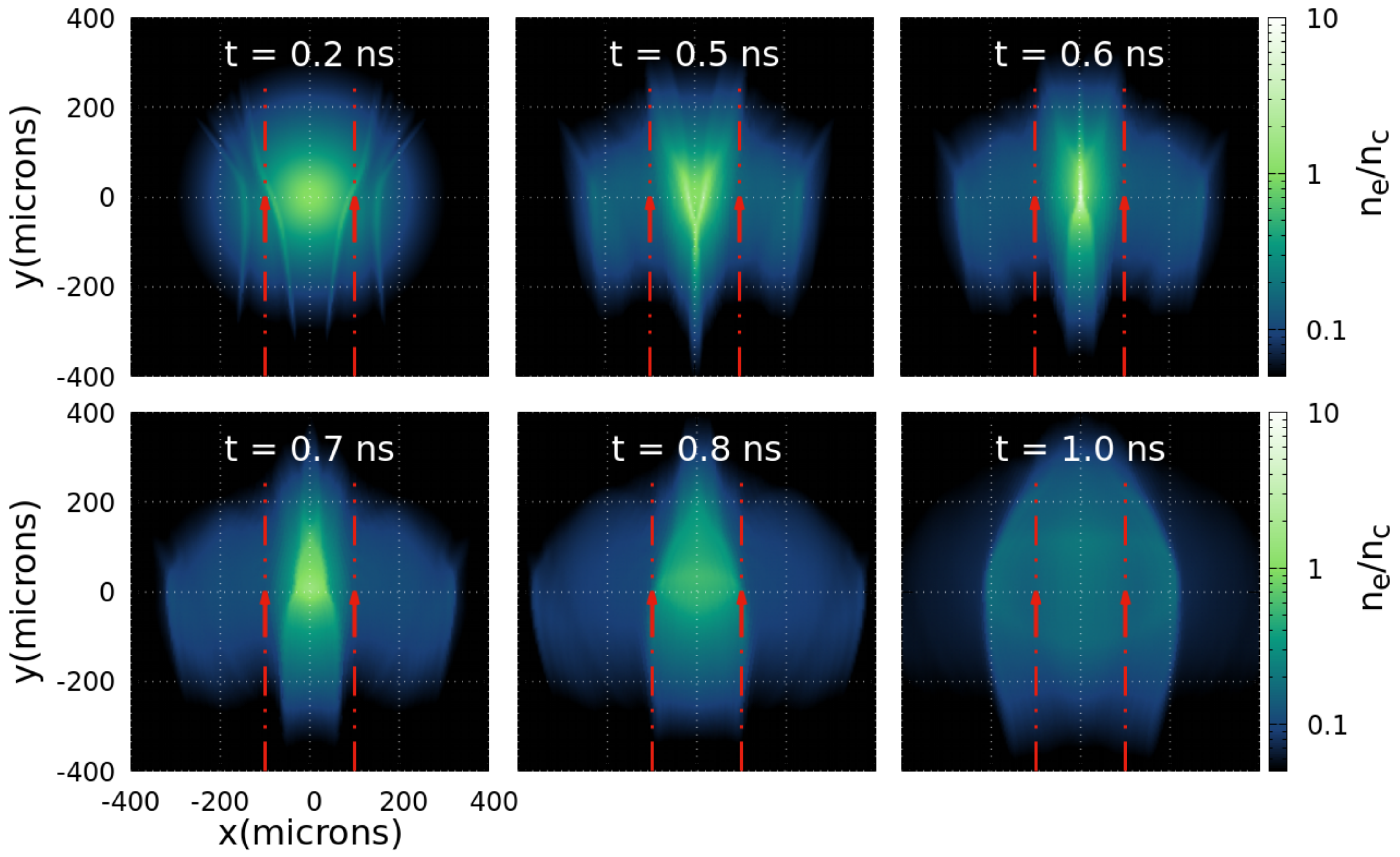}
	\caption{\label{Profil_2D_Dx_100microns}\textit{2D density profiles at different times, for the two beams propagating at $\Delta x$ = $\pm 100$ $\mu$m from the jet center (red arrows). Same laser-plasma parameters than in Fig. \ref{Profil_2D_2faisceaux_fcn_temps}.}}
\end{figure}

Figure \ref{Profil_2D_Dx_100microns} presents a case where the two ns-beams propagate at $\Delta x = \pm 100 \mu$m. Since the laser beams propagate in the core of the gas jet, they undergo a strong refraction, leading to initially very tilted shock fronts. A relatively flat density profile at the collision point is nevertheless generated ($t=0.6$ ns). However, the density reduction in the  wings of the gas jet is less efficient than for $\Delta x = \pm 200$ $\mu$m (Fig. \ref{Profil_2D_2faisceaux_fcn_temps}). The plasma compression is also less efficient, $n \sim 5 n_0$, and lasts for a shorter time. 

The velocity of the BW at a distance $r$ from its origin can be retrieved from Eq. \ref{r_BW}:

\begin{equation}\label{v_BW}
v_{BW}(r) = \frac{2}{2+\alpha} \zeta(\gamma,\alpha)^{(2+\alpha)/2} \left(\frac{E_0}{\rho}\right)^{1/2} r^{-\alpha/2}
\end{equation}

The farther is the BW from its origin, the slower it propagates. In the case of Fig. \ref{Profil_2D_Dx_100microns}, the collision of the BWs occurs close to their starting points, with high velocities ($\sim 200$ $\mu$m/ns). They quickly interpenetrate, and the central thin-plasma quickly expands and decreases in density.

\begin{figure}
	\includegraphics[width=\columnwidth]{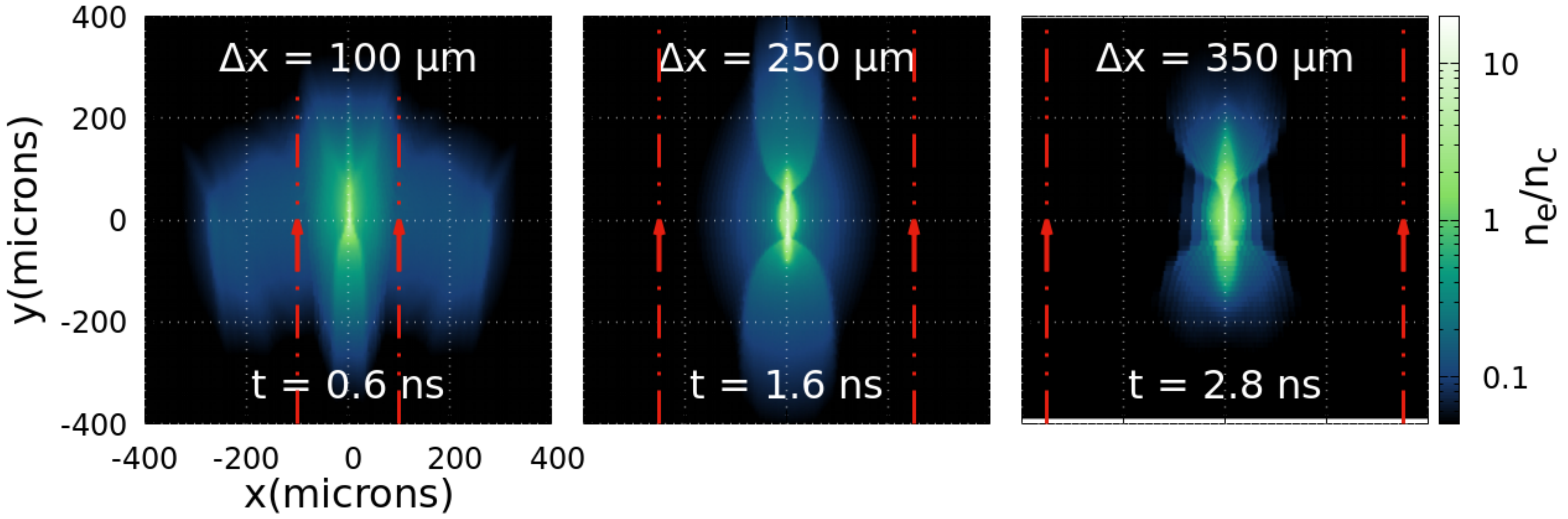}
	\caption{\label{Profil_2D_fcn_distance}\textit{2D density profiles at the collision time of the two BWs, for three distances $\Delta x$ of the ns-beams from the jet center (red arrows). Same laser-plasma parameters than in Fig. \ref{Profil_2D_2faisceaux_fcn_temps}.}}
\end{figure}

\begin{figure}
	\includegraphics[width=\columnwidth]{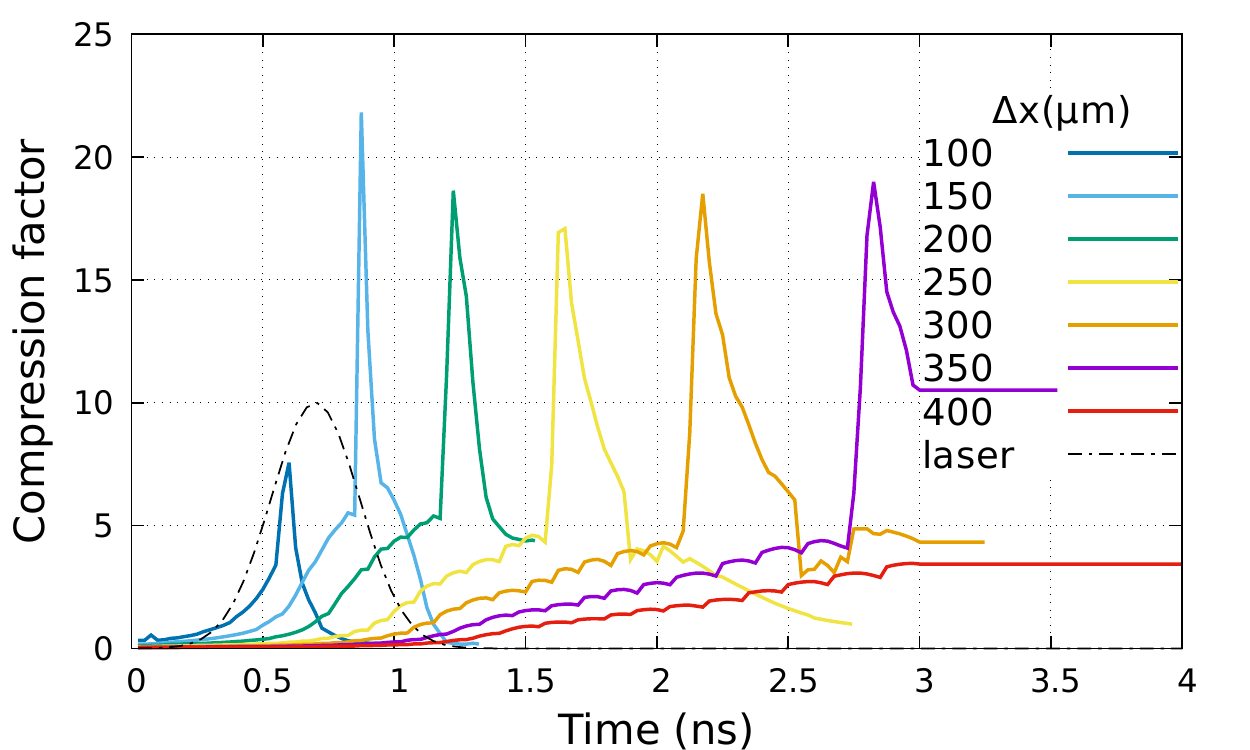}
	\caption{\label{Compression_fcn_temps_et_Dx}\textit{Evolution of the compression factor $n^{front}/n_e$ at $y=0$, during the propagation of the BWs, for different positions $\Delta x$ of the laser beams. The black-dashed curve shows the laser pulse temporal profile. The laser parameters are the same as in Fig. \ref{Profil_2D_2faisceaux_fcn_temps}, the initial density is $n_0/n_c$ = 1.}}
\end{figure}

Figure \ref{Compression_fcn_temps_et_Dx} shows the evolution of the compression factor $n^{front}/n_e$ at $y=0$, during the propagation of the BWs, for different positions $\Delta x$ of the laser beams. For each curve the peak occurs when the two BWs collide (at $x=0$). For small values of $\Delta x$, the maximum compression of the BW before collision is close to the maximum value $(\gamma+1)/(\gamma-1)\sim$ 6 expected for H$_2$. When $\Delta x$ is increased, the beams propagate in a lower density plasma, the energy deposition is lower, resulting in a slightly lower maximum compression of each BW ($\sim$ 4 for $\Delta x$ = 350 $\mu$m). At the BWs collision, the compression factor is multiplied by almost a factor 4 ($n^{front}/n_e \sim 18$). The larger $\Delta x$, the longer the BWs have traveled before their collision, and the slower are their respective velocities (Eq. \ref{v_BW}). This increases the duration of the collision: for a compression factor $>$ 10, it increases from 50 ps for $\Delta x$ = 150 $\mu$m to 180 ps for $\Delta x$ = 300 $\mu$m, and then it remains approximatively constant for $\Delta x$ = 350 $\mu$m. For $\Delta x \ge$ 400 $\mu$m, the energy deposition is too low and the BWs do not collide and are reflected before reaching the jet center.

\begin{figure}
	\includegraphics[width=\columnwidth]{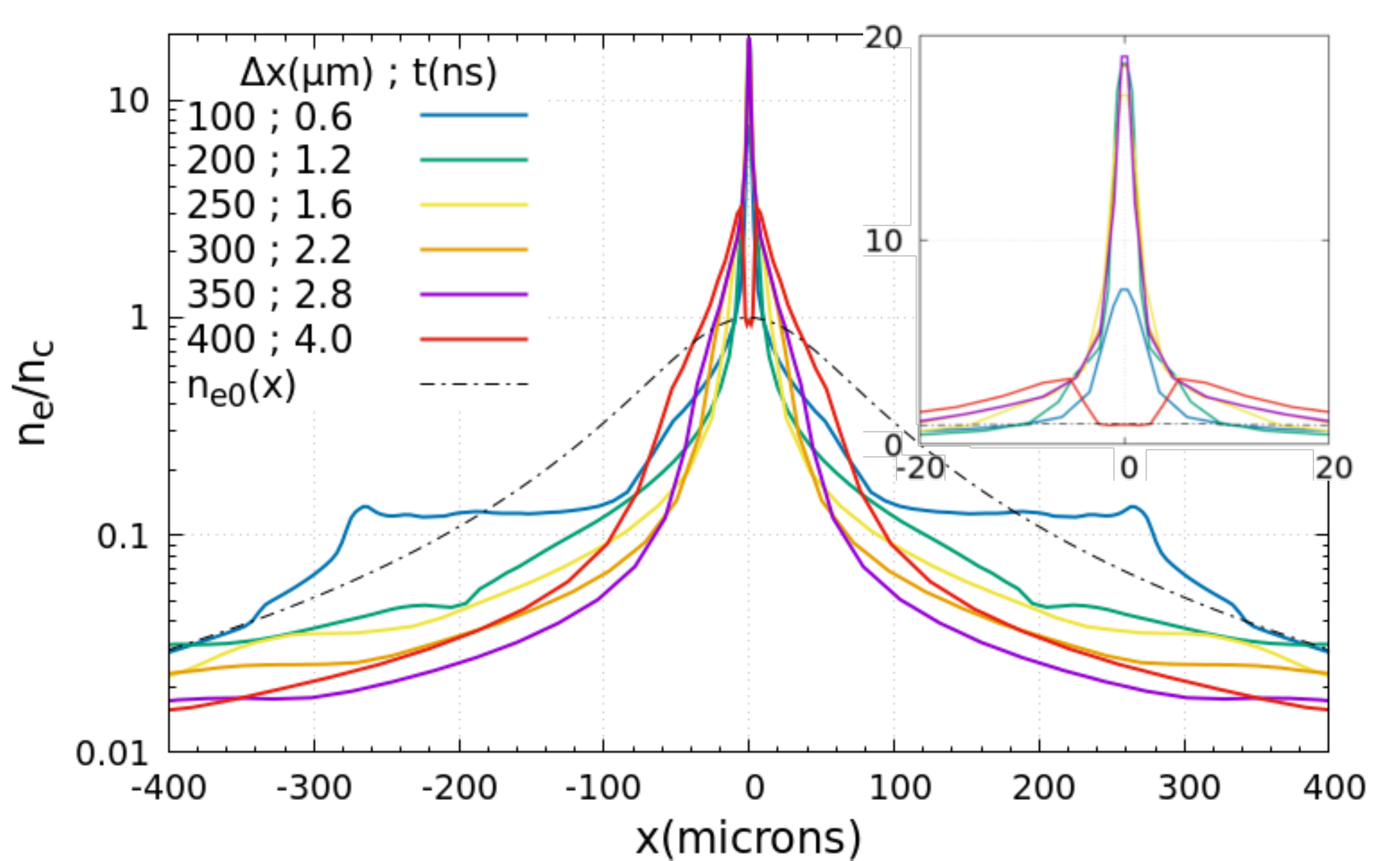}
	\caption{\label{Coupes_2faisceaux_fcn_distance_laser}\textit{Lineouts of the density profiles at $y=0$, at the time of the BW collisions, and for different positions $\Delta x$ of the lasers to the jet center. The black-dashed curve is the initial density profile. The inset in the right-top corner is a zoom of the central region, in linear scale. The laser parameters are the same than in Fig. \ref{2D_profiles_1beam_vs_ne0}, \ref{2D_1beam_late_time}, and \ref{Profil_2D_2faisceaux_fcn_temps}.}}
\end{figure}

Line-outs of the density profiles at the collision time, obtained for different position $\Delta x$ of the ns-beams, are presented in figure \ref{Coupes_2faisceaux_fcn_distance_laser}. In the $\Delta x = 100$ $\mu$m case, the expulsion of the plasma induced by the BW occurs in a dense part of the gas jet, and the time elapsed before the collision is too short to allow an efficient digging of the density profile, resulting in a relatively high density ($\sim 0.1 n_c$) large plateau ($\sim 200$ $\mu$m) in front of the collision zone, and a relatively low final compression ($\sim 7.5$). In contrast, the $\Delta x = 350$  $\mu$m case shows an efficient expulsion of the plasma and compression toward the collision zone. The thin, sharp gradient, transient plasma is denser ($\sim 18 n_0$) and, since the BWs collide at slower velocities, lasts longer. As shown in Fig. \ref{Compression_fcn_temps_et_Dx}, for $\Delta x = 400$ $\mu$m the two BWs never reach the center of the gas jet and are reflected before colliding (the red curve corresponds to the time (4.0 ns) at which the BWs are the closest to the jet center, at later times the BWs motion is reversed).

For $\Delta x \ge$ 200 $\mu$m, as refraction of the beams is weak, changing the initial density $n_0$ (backing pressure of the gas jet) from $n_0/n_c=1$ to $n_0/n_c$ = 1/10 does not modify the shape of the density profile obtained at the BWs collision (Fig. \ref{Coupes_2faisceaux_fcn_distance_laser}), but only its peak amplitude, as previously shown in the case of the tailoring by a single beam (Fig. \ref{2D_profiles_1beam_vs_ne0}).

\begin{figure}
	\includegraphics[width=\columnwidth]{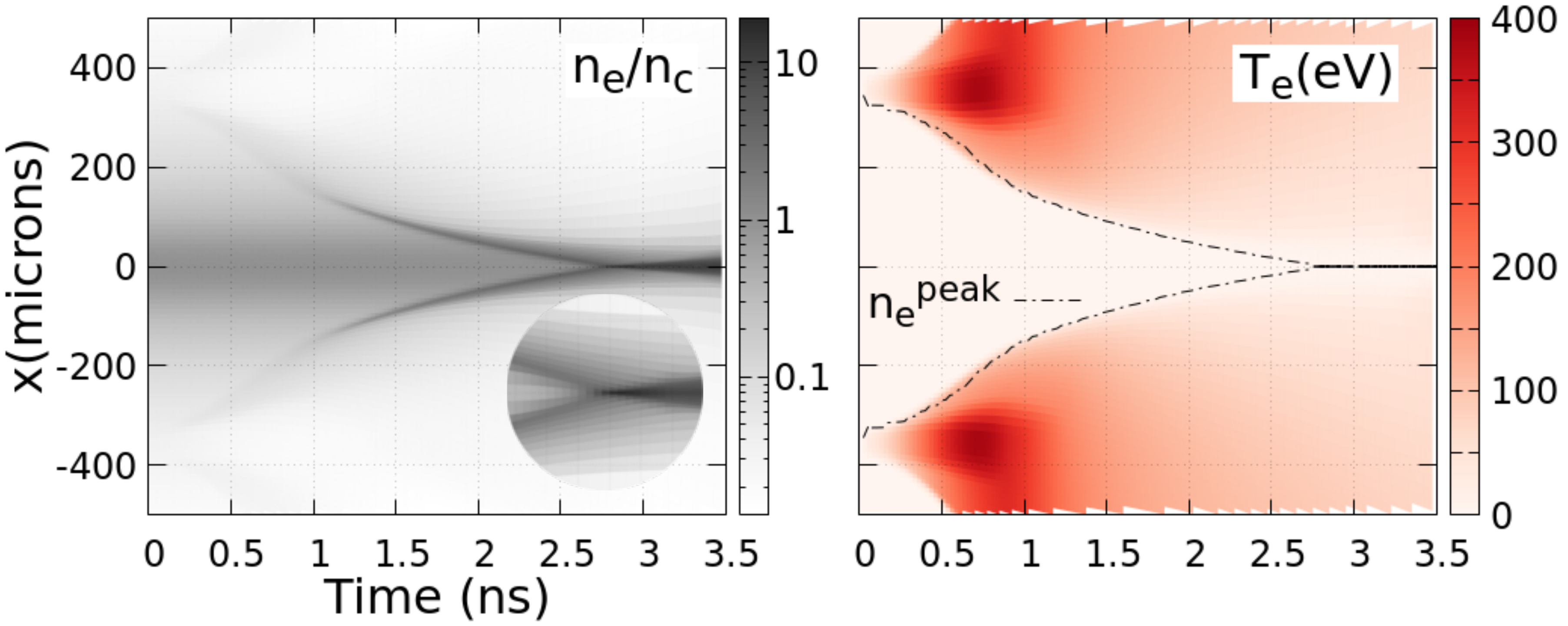}
	\caption{\label{nesurnc_Te_350mic_vs_time}\textit{Evolution of the density (left) and temperature (right) perturbations in the transverse direction (along $x$) at $y$ = 0, for the case $\Delta x$ = 350 $\mu$m of Fig. \ref{Profil_2D_fcn_distance}-\ref{Coupes_2faisceaux_fcn_distance_laser}. The inset in the left graph is a zoom near the collision of the BWs.}}
\end{figure}

\begin{figure}
	\includegraphics[width=\columnwidth]{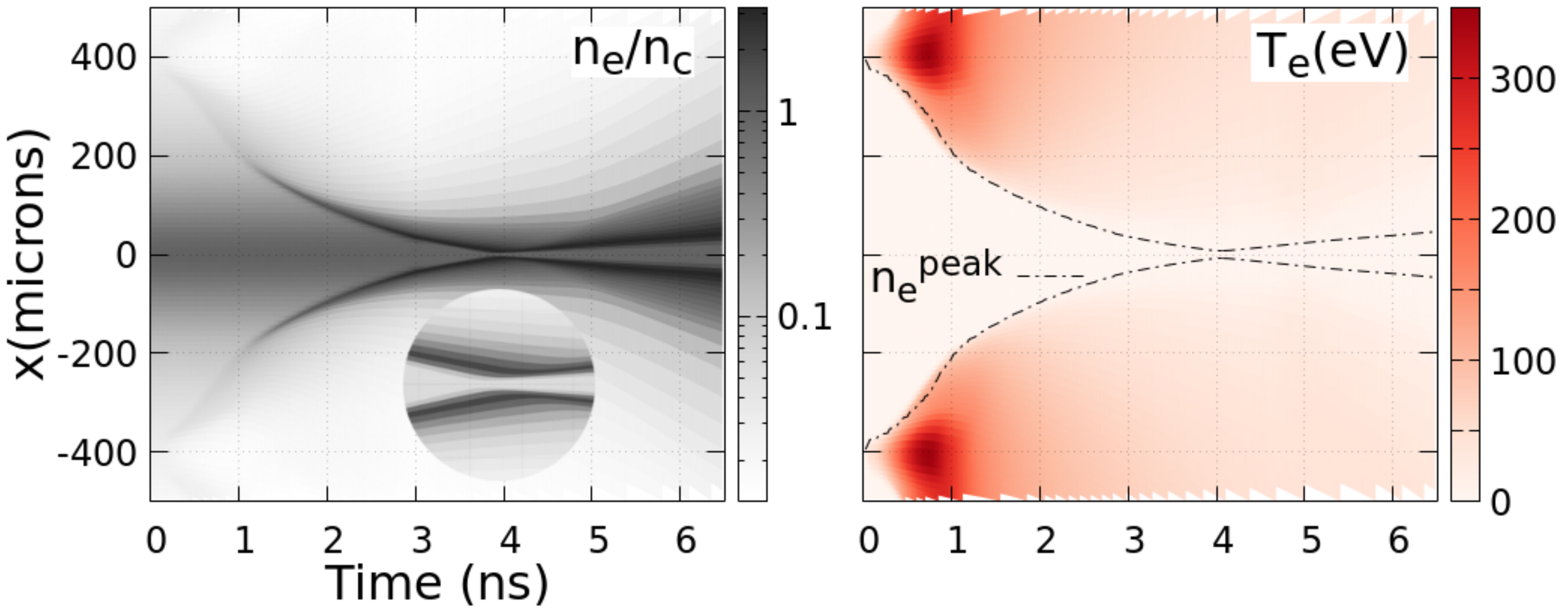}
	\caption{\label{nesurnc_Te_400mic_vs_time}\textit{Same as Fig. \ref{nesurnc_Te_350mic_vs_time} for the case $\Delta x$ = 400 $\mu$m. The inset in the left graph is a zoom showing that the BWs are reflected before reaching the jet center, and thus do not collide/interpenetrate.}}
\end{figure}

The propagation of the density and temperature perturbations along the $x$-axis (at $y$ = 0) is presented in figure \ref{nesurnc_Te_350mic_vs_time} and figure \ref{nesurnc_Te_400mic_vs_time} for the case  for the case $\Delta x$ = 350 $\mu$m and 400 $\mu$m respectively. As shown in Fig. \ref{Compression_fcn_temps_et_Dx}, the two counterpropagating BWs propagate toward $x$ = 0 and heat the plasma at their front. In the case $\Delta x$ = 400 $\mu$m, the two BWs never reach the jet center: close to $t$ = 4 ns they are reflected before colliding. In the case $\Delta x$ = 350 $\mu$m the initial energy in the BWs is large enough to allow them to reach the jet center and collide (at $t \sim$ 2.8 ns). The resulting high-density sharp-gradient low-temperature plasma is preserved during at least 1 ns, suggesting a stagnation process.


\begin{figure}
	\includegraphics[width=\columnwidth]{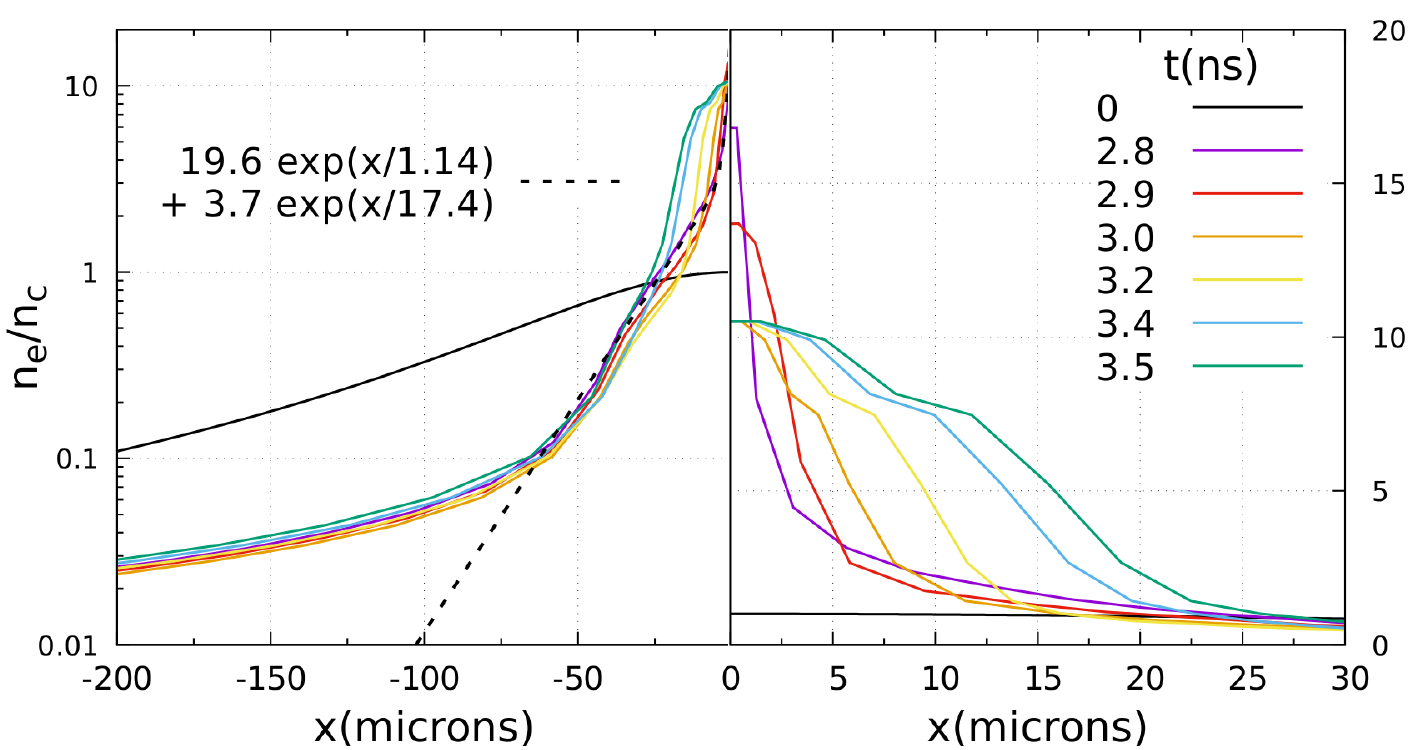}
	\caption{\label{Coupes_temps_long_2beams}\textit{Lineouts of the density profiles at $y=0$, at times after the BWs collision ($t$ = 2.8 ns), for the case $\Delta x$ = 350 $\mu$m of Fig. \ref{Coupes_2faisceaux_fcn_distance_laser}-\ref{nesurnc_Te_350mic_vs_time}. Given the symmetry of the profile, the density is represented in logarithmic scale on the left side, and in linear scale on the right side (with a zoom along the $x$-axis). The black curve is the initial density profile, the black-dashed curve is a fit of the profile at $t$ = 2.8 ns.}}
\end{figure}

Lineouts of the density profiles at different times after the BWs collision are presented in figure \ref{Coupes_temps_long_2beams} for the case $\Delta x$ = 350 $\mu$m. At the collision time, $t=2.8$ ns, the plasma is very thin, FWHM $\sim$ 5 $\mu$m. The black-dashed line shows a fit of this profile, indicating that a very small gradient length ($<17.4$ $\mu$m) is obtained on more than 2 orders of magnitude. After 100 ps, the plasma is two times larger while its density has only decreased by 20-25 $\%$. From $t$ = 3 ns to 3.6 ns, the maximum compression factor stagnates at $\sim$ 10, while the plasma width increases by more than a factor 6. As discussed at the end of section \ref{Single_beam_tailoring}, this is a consequence of the bow shape evolution that tends to transversally ($y$-axis) compress the plasma from the gas jet edge toward the propagation axis ($x$-axis). This can be been observed in Fig. \ref{Profil_2D_2faisceaux_fcn_temps} at times after the collision, $t>$ 1.2 ns. Despite this increase in width, the plasma wings remain at a relatively low density, preserving the high contrast of the plasma profile (left scale in Fig. \ref{Coupes_temps_long_2beams}). By choosing the time after the BWs collision, it is thus possible to produce a high-density steep-gradient plasma of adjustable thickness.

\begin{figure}
	\includegraphics[width=\columnwidth]{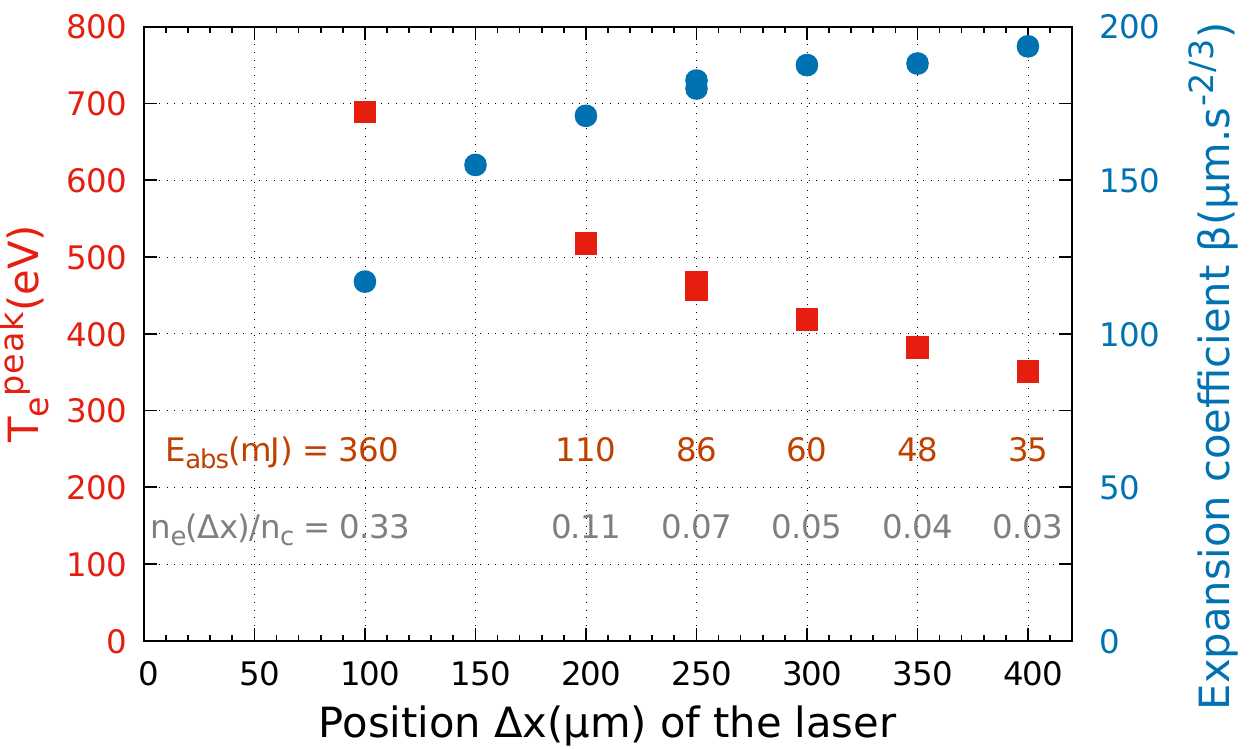}
	\caption{\label{Te_expansion_vs_distance}\textit{Dependence on the position $\Delta x$ of the laser beams. Red, left scale: peak value of $T_e$ (reached on the laser axis, at the pulse maximum). Blue, right scale: factor $\beta$ of the radial expansion $r(t) = \beta t^{2/3}$. For each set of ($T_e$,$\beta$) points are reported, in grey values the absorbed laser energy $E_{abs}$ in the plasma, and in dark-orange values the plasma density at the laser position. The laser parameters are the same as in figures \ref{2D_profiles_1beam_vs_ne0}, \ref{2D_1beam_late_time}, and \ref{Profil_2D_2faisceaux_fcn_temps}, the initial density is $n_0/n_c$ = 1.}}
\end{figure}

The dependence of the initial plasma temperature ($T_e$) and of the factor $\beta$ that governs the ensued BW expansion (velocity) is presented in figure \ref{Te_expansion_vs_distance} as a function of $\Delta x$. At large $\Delta x$, the beams propagate in the low density wing of the gas jet, resulting in a low energy deposition $E_{abs}$ and a low heating. However, since $\beta$ depends on the ratio $E_0/\rho$ (Eq. \ref{r_BW} and \ref{v_BW}), it increases with $\Delta x$, up to a position where the plasma density is so low that the energy deposition becomes too small to excite a strong hydrodynamic shock. In the present case, the inflexion point is at $\Delta x \sim$ 300 $\mu$m, which is more than 4 times the radius of the gas jet. From $\Delta x$ = 200 to 400 $\mu$m, the expansion coefficient $\beta$ varies by less than 10 $\%$. The plasma dynamics at the collision point depends on the relative velocity of the two BWs (see below). To tend toward a stagnation regime, the velocity of the BWs needs to be low. At the collision point it is given by Eq. \ref{v_BW}, with $r = \Delta x$. The term $E_0/\rho$ is proportional to $\beta$, which weakly depends on $\Delta x$, but also weakly on the laser energy $E_L$ (see discussion above). The most efficient way to reduce the relative velocity of the BWs is thus through the term $r^{-\alpha/2}$, that is by increasing the distance before collision, and thus $\Delta x$. This is possible up to a value of $\Delta x$ for which the initial energy of the BW becomes too small to compensate the slowing down induced by the density gradient of the background plasma (gas jet profile). This is what happens in the case $\Delta x$ = 400 $\mu$m where the initial energy of the BWs is low, the maximum compression of each BW does not exceed $\sim$ 3, and their motion is stopped before reaching the plasma center, preventing their collision or interpenetration.

\subsubsection{Influence of the initial density profile of the gas jet}
\begin{figure}
	\includegraphics[width=\columnwidth]{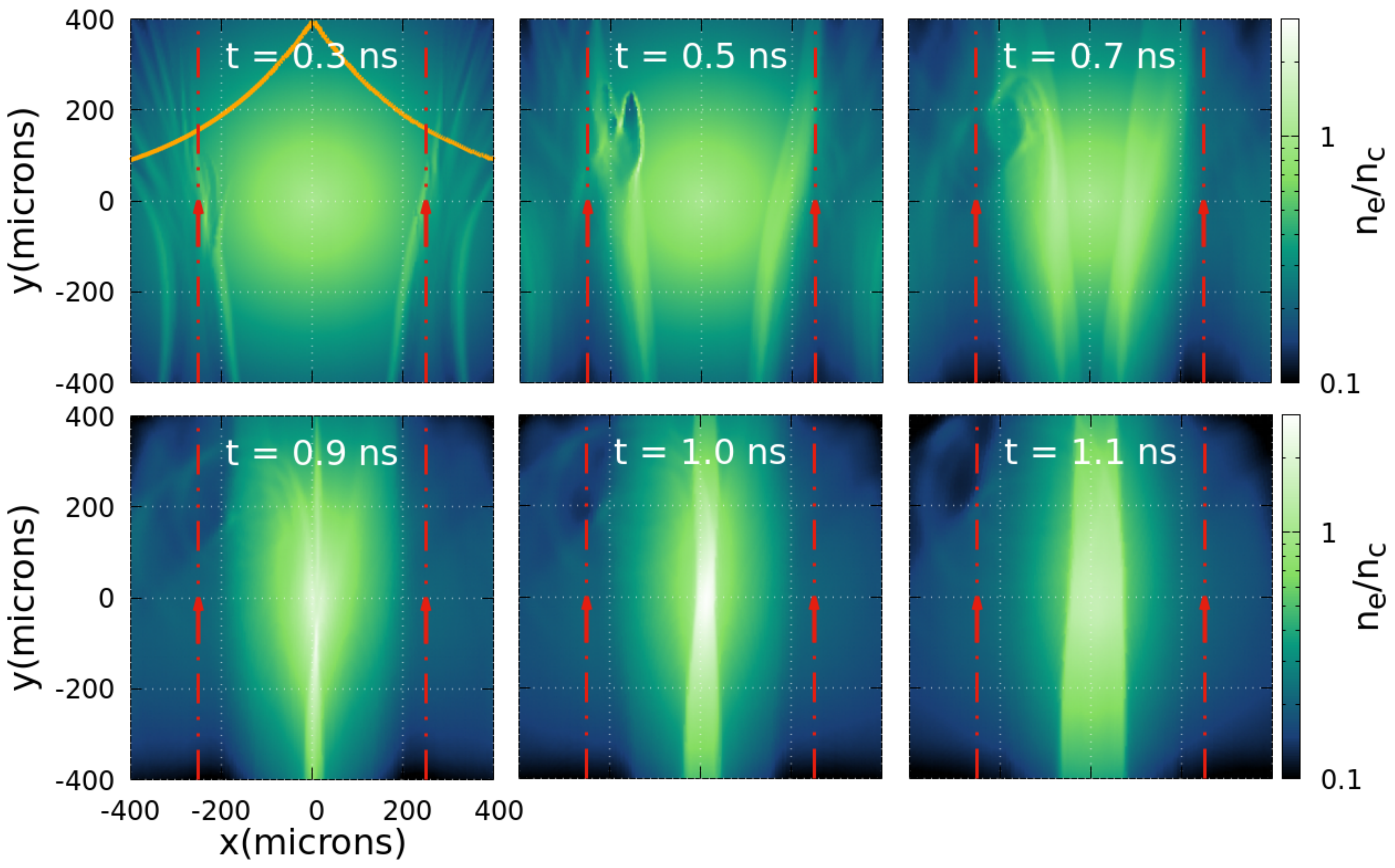}
	\caption{\label{Profil_2D_Laval2}\textit{2D density profiles at different times, for an initially broad gas jet (FWHM $\sim 400 \mu$m). The ns-beams propagate at $\Delta x$ = $\pm 250$ $\mu$m, and $n_0/n_c$ = 1.}}
\end{figure}

\begin{figure}
	\includegraphics[width=\columnwidth]{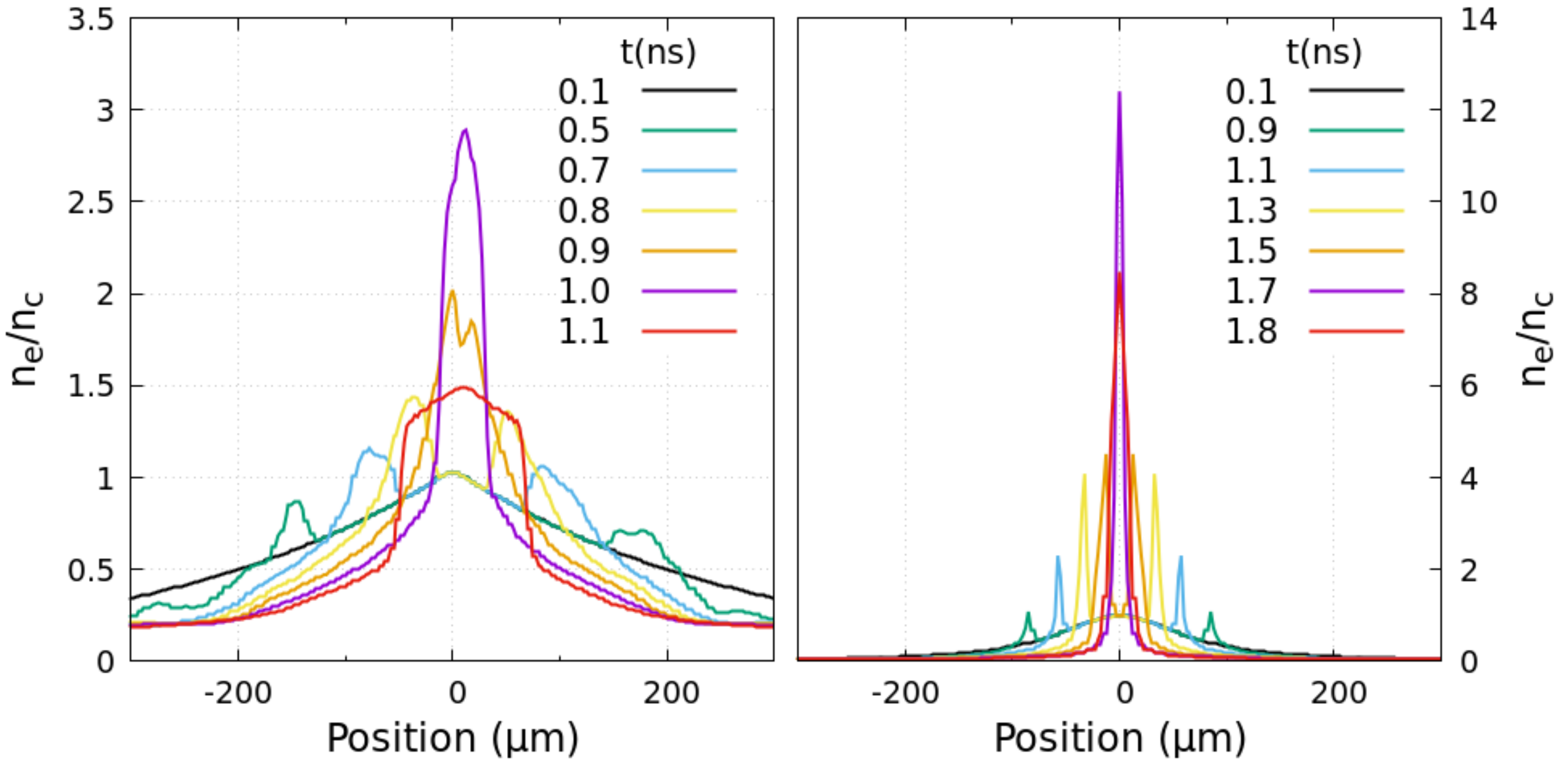}
	\caption{\label{Coupes_densite_Laval1_Laval2}\textit{Line-outs of the density profiles at $y=0$, at different times, and for two initial density profiles: FWHM = 400 $\mu$m (left), 140 $\mu$m (right). The lasers propagate at $\Delta x$ = $\pm 250$ $\mu$m, the initial peak density is $n_0/n_c$ = 1.}}
\end{figure}

The initial density profile of the gas jet is also a crucial parameter. This is illustrated in figure \ref{Profil_2D_Laval2} where the width of the gas jet is $\sim$ 400 $\mu$m (FWHM), instead of $140 \mu$m in the previously presented cases. A comparison of the evolution of the density profile (at $y=0$) for two types of initial profile is shown in figure \ref{Coupes_densite_Laval1_Laval2}. With the broader profile, the higher density along the laser axis leads to a $\sim$ 2 times higher plasma temperature ($T_e \sim 1$ keV), and higher velocities of the BWs. In addition, each laser beam independently undergoes a strong refraction and filamentation, leading to asymmetric BWs. The consequences are a lower compression in each BW and a less efficient collision, producing a broader and asymmetric plasma of lower density, $\sim 3 n_0$ (Fig. \ref{Coupes_densite_Laval1_Laval2}-right), a factor $\sim 4$ compared with the case of the initially thinner gas jet (Fig. \ref{Coupes_densite_Laval1_Laval2}-left). From Fig. \ref{Profil_2D_Laval2} it follows that due to the smoother density profile associated with this larger width, the BWs do not develop a bow shape.

\subsubsection{Influence of the type of gas}
\begin{figure}
	\includegraphics[width=\columnwidth]{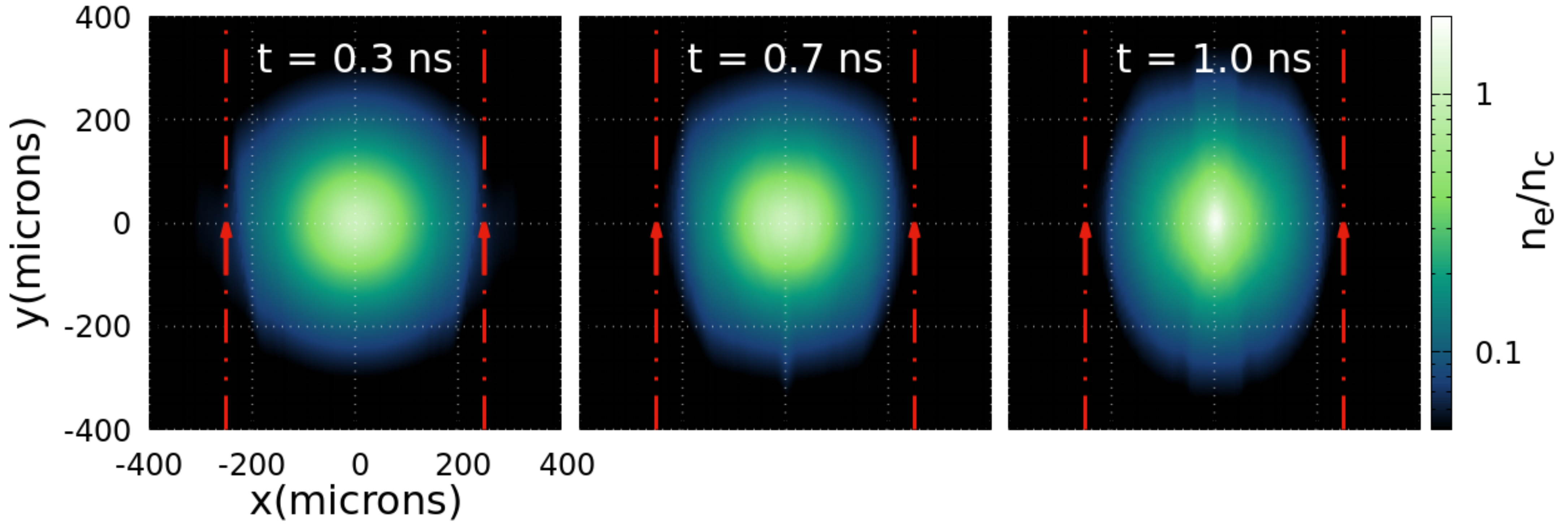}
	\caption{\label{Profil_2D_He}\textit{2D density profiles at different times, for a Helium plasma. The initial density is $n_0/n_c$ = 1, the spatial profile is the same than in Fig. \ref{1faisceau_2D} to \ref{Coupes_2faisceaux_fcn_distance_laser}. The two 25 J ns-beams propagate at $\Delta x$ = $\pm 250$ $\mu$m from the jet center (red arrows).}}
\end{figure}

The initial hydrodynamic shock and the BW properties also depend on the type of gas, through the heat capacity ratio $\gamma$, the atomic number $Z$ and the mass $m_i$ of the ions that define the bremsstrahlung absorption, the hydrodynamic compression, and the BW velocity (Eq. \ref{v_BW})\cite{Sedov_book}. Figure \ref{Profil_2D_He} presents 2D density profiles obtained using a Helium gas ($\gamma$ = 5/3). The other parameters are the same as for simulation in Fig. \ref{Coupes_densite_Laval1_Laval2}-left: the beams propagate at $\Delta = \pm 250$ $\mu$m, and the gas jet has a narrow profile (FWHM = 140 $\mu$m) with an initial peak density of $n_0/n_c$ = 1. Compared to hydrogen, the larger $Z$ of Helium increases the laser absorption: $E_{abs}$ = 250 mJ and $T_e \sim$ 800 eV, while $E_{abs}$ = 140 mJ and $T_e \sim$ 500 eV for H$_2$. Nevertheless, the larger mass density $\rho$ (for a given $n_e$) and the lower $\gamma$ lead to weaker and broader shock fronts. The BWs collision produces much lower peak density, $\sim 1.8 n_0$, with much weaker gradients.

The conclusions on the efficiency of the plasma expulsion versus $\Delta x$, the initial density profile, or the type of gas also apply to the plasma tailoring by a single beam discussed in section \ref{Single_beam_tailoring}.

\subsection{Blast-wave collision in a three-dimensional geometry}
Using planar TROLL simulations, we have explored the physics and the key parameters involved in the laser-produced hydrodynamic shocks, and the ensued plasma tailoring by the BWs. However, even if a supersonic gas jet is relatively homogeneous along the gas flow  ($z$ in our referential), in the present scheme the laser beams are perpendicular to the jet and have a cylindrical geometry, making the propagation and collision of the BWs a three-dimensional process (see Fig. \ref{Principe}). In particular, the velocities and the shape of the colliding BW fronts might affect the efficiency of the plasma tailoring, in terms of maximum compression (density), size, and uniformity. To estimate such effects, three-dimensional simulations were performed. Even if TROLL can also be run in a 3D geometry, for practical reasons, such as avoiding unwanted mesh deformations - and  guided by the relative simplicity of the set up, we use the 3D Eulerian hydrodynamic code HERA \cite{HERA, HERA_2}. This code is well suited for multi-dimensional hydrodynamic shock simulations, but is not a radiation hydrodynamic code. We mimic the laser energy deposition by imposing, on the background electron temperature, an initial localized electron temperature perturbation in a form a cylinder, corresponding to the laser propagation volume. For simplicity, the electron temperature has only a radial dependence corresponding to the laser focal spot used in TROLL. This initial condition is sufficient to launch a BW. Adding a temporal shape to this electron temperature profile (reproducing the laser pulse) gives similar results. The gas jet has a cylindrical geometry, with a constant profile along the $z$-axis, and the same radial profile in the $x-y$ plane as in TROLL simulations (FWHM = 140 $\mu$m).

\begin{figure}
	\includegraphics[width=\columnwidth]{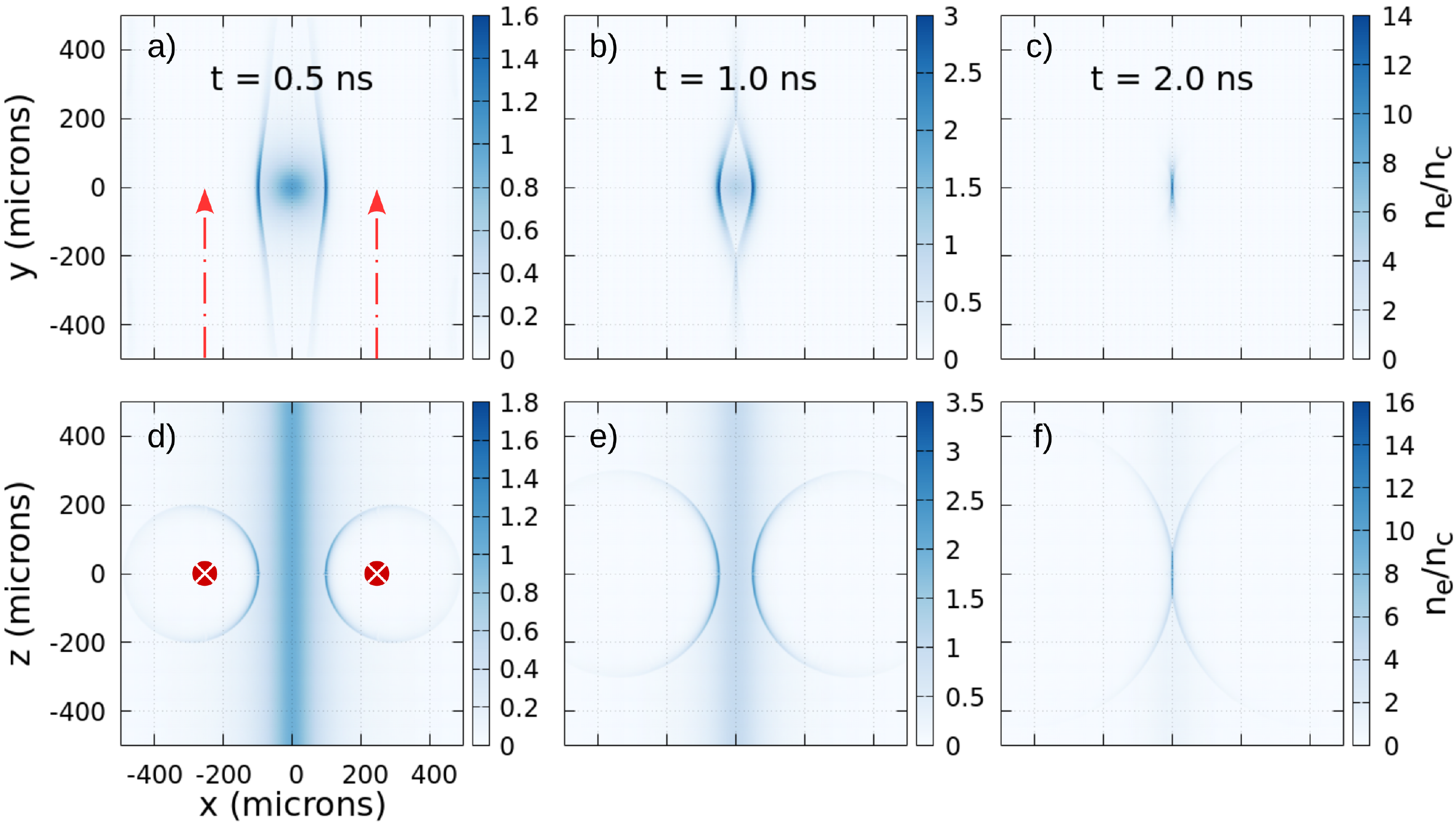}
	\caption{\label{Profils_3D}\textit{2D projections of the plasma density extracted from a 3D HERA simulation. Graphs a)-c) are the density profiles in the $x$-$y$ plane at $z=0$, and d)-f) the profiles in the $x$-$z$ plane at $y=0$. The initial heating axis are at $\Delta x$ = $\pm 250$ $\mu$m and $z=0$. The initial peak density is $n_0/n_c$ = 1.}}
\end{figure}

Two-dimensional maps of the plasma density extracted from a 3D HERA simulation are presented in figure \ref{Profils_3D} at different times, for the case $\Delta x$ = 250 $\mu$m - $n_0/n_c$ = 1 presented in Fig. \ref{Profil_2D_fcn_distance}. Figures \ref{Profils_3D}-a)-c) are the maps in the $x-y$ plane. They are very similar to the TROLL 2D-simulations, showing the same bow shape distortion, BW collision, resulting in a very similar final compression ($n/n_0 \sim 14$). The collision occurs at $t$ = 2 ns, slightly later than in the TROLL simulation (1.6 ns). Figures \ref{Profils_3D}-d)-f) show the evolution of the BWs in the perpendicular plane $x-z$. The expansion has a cylindrical shape. However, while the expansion starts from the laser axis, the propagation along the $x$-axis quickly shows (Fig. \ref{Profils_3D}-d)) a non-symmetric behavior: the velocity of the BW decreases when it encounters a larger density (toward the jet center), while it increases when the density becomes lower (away from the jet center). This effect also contributes to the bow shape observed in the $x-y$ plane.

\begin{figure}
	\includegraphics[width=\columnwidth]{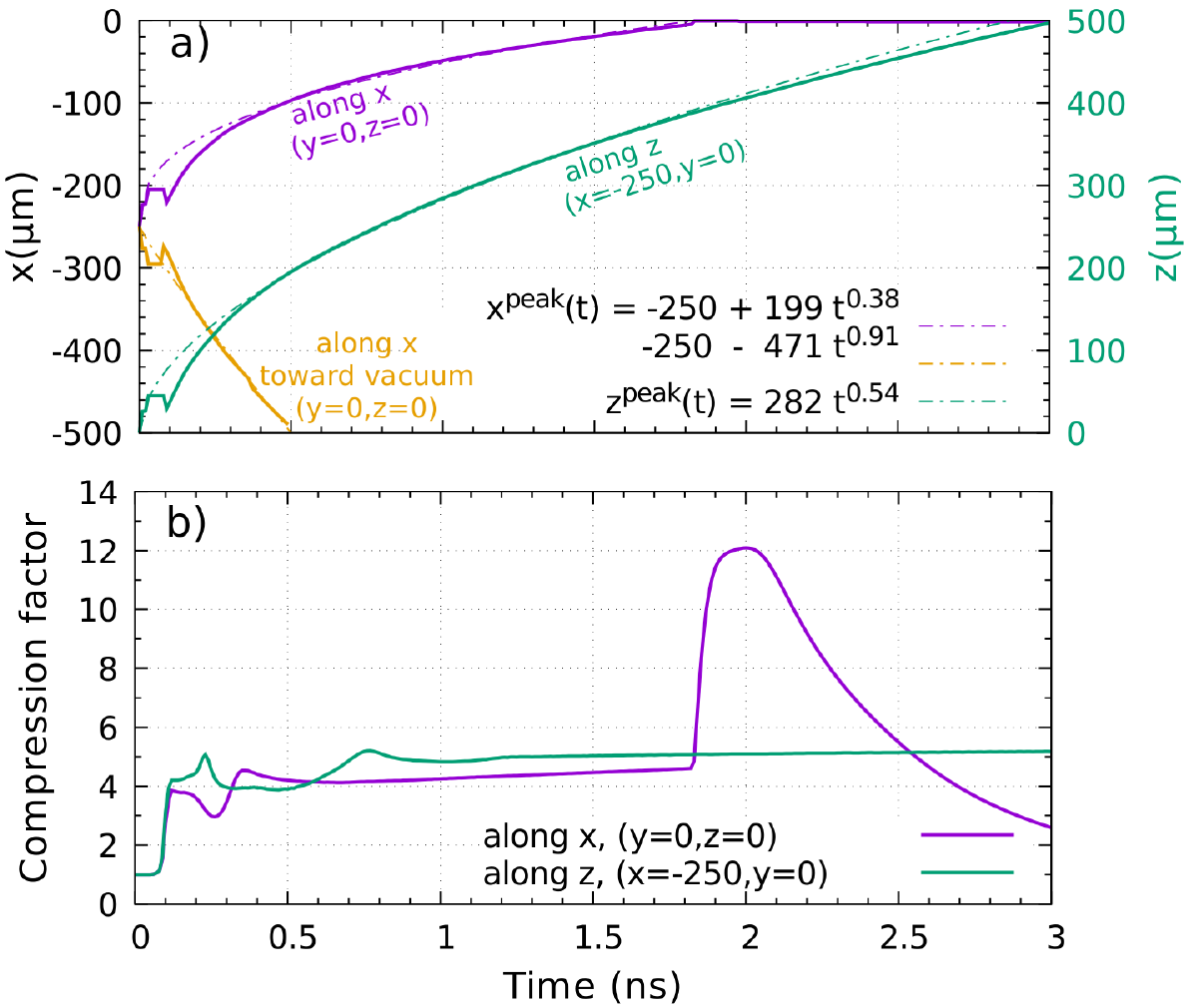}
	\caption{\label{Amplitude_position_vs_time_HERA}\textit{Evolution of a) the position and b) the amplitude of the left-side BW (laser at $x$ = -250 $\mu$m) of Fig. \ref{Profils_3D}, along the $x$ and the $z$ axis. The left scale is for the propagation along the $x$-axis, the right scale for the $z$-axis. The dash-dot lines are fits.}}
\end{figure}

Evolution of the position and amplitude of the left BW along the $x$ and the $z$ axes is presented in figure \ref{Amplitude_position_vs_time_HERA}-a), while the associated compression factors are shown in Fig.\ref{Amplitude_position_vs_time_HERA}-b). Along the $z$-axis (at $x$ = 250 $\mu$m, laser position) the plasma density is constant, and the BW follows a cylindrical expansion ($\propto t^{1/2}$, green curve), as expected. Its compression reaches $n/n_0 \sim 5$, close to the maximum value expected for a hydrogen plasma. On the $x$-axis, the density gradient modifies the BW expansion. As already observed in Fig. \ref{1faisceau_coupes_fcn_temps}, the shock front that travels toward vacuum quickly expands, while the one that propagates in a plasma of increasing density (gas jet center) slows down\cite{Laumbach,Ostriker}, with a time dependency very close to a spherical expansion ($\propto t^{2/5}$). The compression factor is $\sim$ 4-5, and reaches $\sim$ 12.5 at the BWs collision, slightly smaller than from 2D-TROLL simulation (see Fig. \ref{Compression_fcn_temps_et_Dx}, $n/n_c \sim$ 17), but with a slightly longer duration.

The 3D HERA simulations are in good qualitative and quantitative agreement with the 2D TROLL simulations, and confirm that the collision of the two BWs leads to a thin, high-density sharp-edge plasma with a relatively high density contrast.

We note here that, while the velocity of a shock wave can be used to characterize the flow of a gas\cite{Dukhovski}, in all the present simulations, the velocity of the gas flow is not taken into account. This is justified in our conditions since the velocity of the BW is $\sim$ 100 times faster than the supersonic gas flow ($\sim$ 1-2 km/s), so that the modification of the BW structure by the supersonic flow is negligible \cite{Kaganovich}.

\subsection{Kinetic effects in the collision of the blast waves}
The 3D HERA simulations confirm that the present tailoring scheme allows producing thin and dense transient plasmas with sharp density gradients. The 2D TROLL simulations also indicated that increasing the laser distance $\Delta x$ tends to reduce the velocity of the BWs at the collision point, increasing the resulting peak density, as well as generating a long-lasting high-density plasma after the beginning of the collision (Fig. \ref{Compression_fcn_temps_et_Dx}, $\Delta x$ = 350 $\mu$m, $t>3$  ns). This could be the signature of a stagnation regime, where the interaction of the BWs is dominated by collisions between ionic species from each of the opposing plasmas. Such a regime occurs when the ion–ion mean-free path, $\lambda_{ii}$, is of the order or smaller than the typical dimensions of the system, the characteristic length of the plasma density gradients $L = n/\nabla n$. When $\lambda_{ii} >> L$ the two plasmas interpenetrate, the collision is short and the resulting high-density plasma has a short lifetime. In opposite, the regime of "hard stagnation" where $\lambda_{ii}<<L$, no interpenetration occurs and the two plasmas will decelerate rapidly through ion-ion collisions \cite{Rambo,Dardis}.

The ion–ion mean-free path of two counterstreaming ion beams is:

\begin{equation}\label{lambda_ii}
\lambda_{ii} = \left( \frac{4 \pi \epsilon_0^2}{e^4}\right) \frac{m_i^2 v_{12}^4}{Z^4 n_i \Lambda_{12}}
\end{equation}

where $\epsilon_0$ is the vacuum permittivity, $e$ the electron charge, $n_i = n_e/Z$ the ion density of a single ion beam, $v_{12}$ the relative speed of the two ion beams, and $\Lambda_{12}$ the Coulomb logarithm for counterstreaming ions in the presence of warm electrons (see Appendix B in \cite{Merritt} for the expression of $\Lambda_{12}$).

The relative speed is twice the velocity of a single BW at the collision point, $v_{12} = 2v_{BW}($x=0$)$. For $\Delta x$ = 200-350 $\mu$m, the 2D TROLL simulations give $v_{12} \sim$ 120-220 $\mu$m/ns. Since the gas jet(plasma) has not been heated before the arrival of the BW, the electron temperature at the density front is very low, typically $T_e \sim$ 5-20 eV. Just before the collision, the two BWs have a density peak of the order of $(\gamma+1)/(\gamma-1) \sim$ 5-6 $n_0$. With $n_0$ between $n_c/10$ and $n_c$, this leads to $\lambda_{ii}$ between 3 and 400 nm, smaller than the gradient length of the system  ($L>$ few microns). The collision is thus expected to lead to a stagnation phase, where the kinetic energy of the flows is progressively transformed into thermal energy. The regime of stagnation could be of interest to increase the lifetime of the transient high density plasma. In case of hard stagnation, the peak density might also be higher. To evaluate the relevance of hydrodynamic single-fluid simulations and consequently the post-collision phase, we performed kinetic simulations using the ionic mono-dimensional Fokker-Planck code FPion \cite{FPion,FPion_2,FPion_3} (1D in space, 3D in velocities). It is a hybrid code where the electrons are treated as a fluid. The initial conditions in density and temperatures are retrieved from TROLL simulations at a time where the BWs are already generated, near the plasma center and before their collision.

\begin{figure}
	\includegraphics[width=\columnwidth]{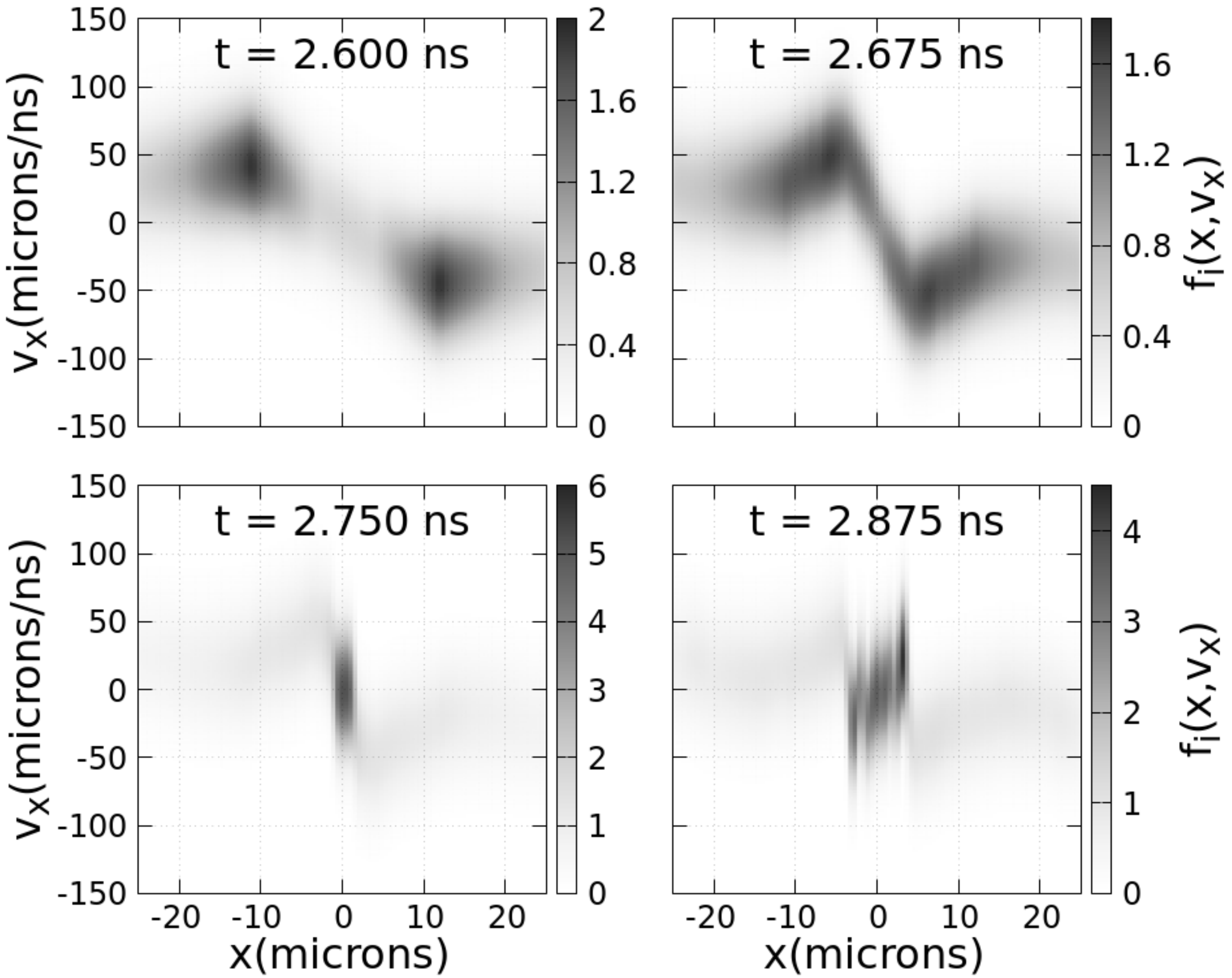}
	\caption{\label{FPion_Dx350_nc}\textit{Ion velocity distribution $f_i(x,v_x)$ from FPion simulations, in the case $\Delta x$ = $\pm 350$ $\mu$m and $n_0$ = $n_c$. The distribution is expressed in unit $n_c/v_{T_i}^3$, where $v_{T_i}$ is the ion thermal velocity for $T_i$ = 10 eV. The simulation starts using the TROLL outputs at $t$ = 2.60 ns.}}
\end{figure}

Figure \ref{FPion_Dx350_nc} presents the evolution of the ion velocity distributions of the two BWs in the collision region, in the case $\Delta x$ = 350 $\mu$m and $n_{0}$ = $n_c$ (figures. \ref{Profil_2D_fcn_distance}-\ref{nesurnc_Te_350mic_vs_time}, \ref{Coupes_temps_long_2beams}). Before interacting ($t$ = 2.6 ns), the BWs fronts have Maxwellian distributions ($v_{T_i} \sim$  31 $\mu$m/ns) with opposite central velocities of $|v_x| \sim$ 50 $\mu$m/ns. After 150 ps ($t$ = 2.75 ns) the two BWs have propagated $\sim$ 12 $\mu$m and reached the plasma center, where they collide. After almost the same amount of time (125 ps, $t$ = 2.875 ns), the fronts of the BWs have covered only $\sim$ 4 $\mu$m, with velocity distributions that are still Gaussian-like, but with central velocities significantly reduced by the collision ($< 25$ $\mu$m/ns). The collision has lead the plasma in a phase of stagnation, as expected, since before collision $\lambda_{ii} \sim$ 2 nm, i.e. much smaller than the characteristic length of the plasma density gradients.

\begin{figure}
	\includegraphics[width=\columnwidth]{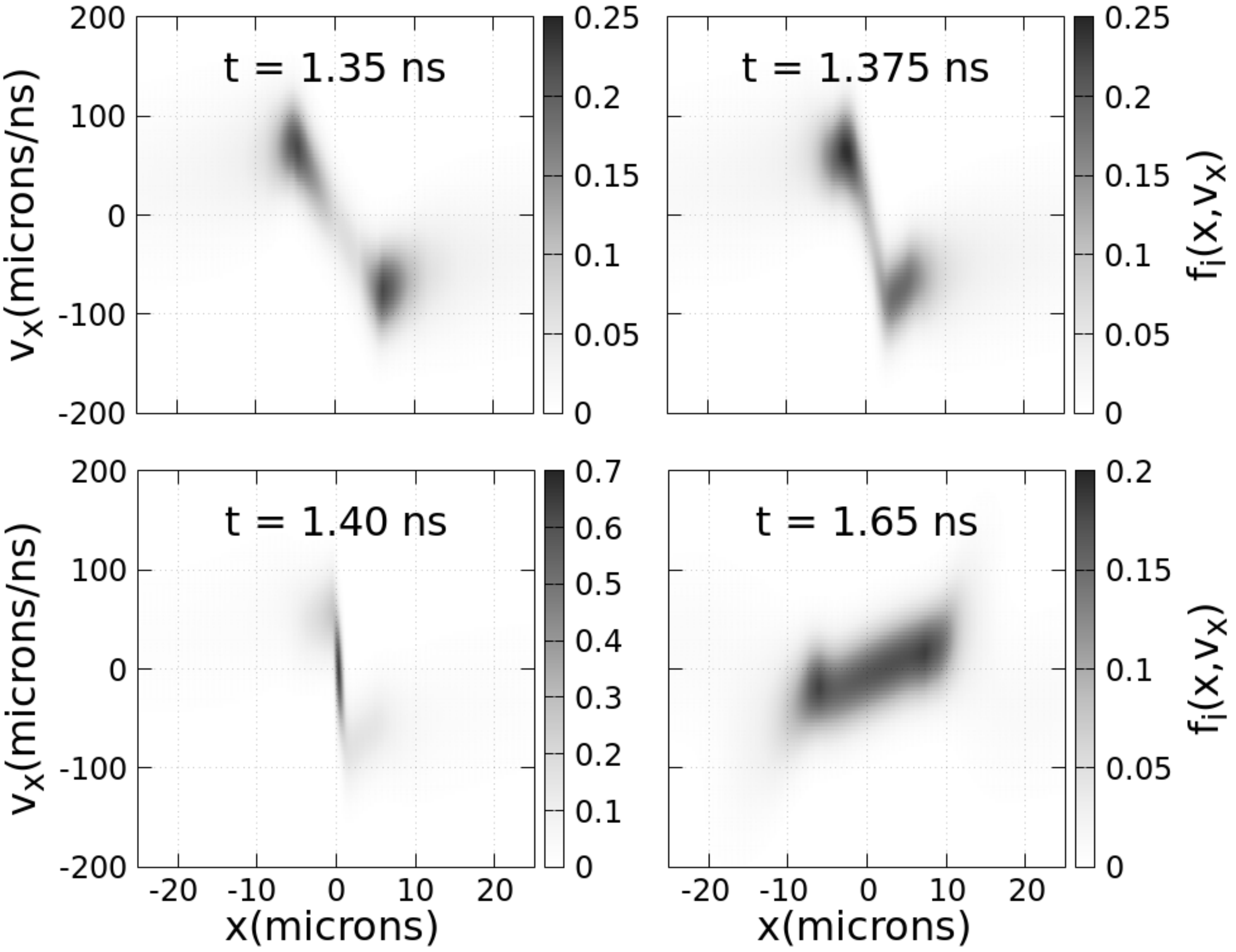}
	\caption{\label{FPion_Dx200_ncsur10}\textit{Ion velocity distribution from FPion simulations, in the case $\Delta x$ = $\pm 200$ $\mu$m and $n_0$ = $n_c/10$. The simulation starts using the TROLL outputs at $t$ = 1.35 ns.}}
\end{figure}

Another typical example of the evolution of the ion velocity distributions is given in figure \ref{FPion_Dx200_ncsur10}, at lower density, $n_0$ = $n_c/10$, and for $\Delta x$ = 200, resulting in larger initial velocities, $|v_x| \sim$ 75 $\mu$m/ns, at only $\sim$ 6 $\mu$m from the plasma center (Fig. \ref{FPion_Dx200_ncsur10} - $t$ = 1.35 ns). After 50 ps ($t$ = 1.40 ns) the BWs collide. However, it takes 200 ps ($t$ = 1.65 ns) for the plasma fronts to cover an equivalent distance (6 $\mu$m), and the average velocity is significantly reduced ($< 20$ $\mu$m/ns). Here again, the plasma is in a phase of stagnation, as expected from the ion-ion mean free path: $\lambda_{ii} \sim$ 80 nm.

The evolution of density profile obtained from FPion is also in good agreement with the TROLL results. This hybrid approach (kinetic for the ions, fluid for the electrons) validates, in our conditions, the single fluid treatment used in TROLL simulations, and confirms that for large values of $\Delta x$ and/or low initial density, the collision of the two BWs could lead to a plasma in a stagnation phase, that increases the lifetime of the transient thin, high-density and sharp-gradient plasma.

\section{Application to proton acceleration}
To study the acceleration process under different scenarios of plasma shaping, particle-in-cell (PIC) simulations were performed using the open-source code Smilei \cite{Smilei}. The 2D TROLL hydrodynamics simulations at a chosen time of the plasma evolution are used as input parameters: the spatial profiles of the electron and ion densities and temperatures are interpolated on a cartesian domain of size 1000 $\mu$m ($x$-axis) $\times$ 60 $\mu$m ($y$-axis), in cells of 60 nm $\times$ 75 nm respectively. Each cell contains 50 macro-particles, corresponding to a total number of $10^9$ particles. The laser driving the proton acceleration has a wavelength $\lambda_0$ = 1 $\mu$m, a Gaussian temporal profile of 1 ps duration (FHWM), and a Gaussian transverse profile with a FWHM = 10 $\mu$m. Its polarization is linear along the $y$-axis. The maximum normalized laser vector potential is set to $a_0$ = 0.6, corresponding to an intensity of $5\times 10^{19}$ W/cm$^2$. The vacuum focus is at the center of the simulation box, which is also the position of the peak of the plasma density profile ($x=0$ and $y$=0 in the above TROLL simulations). The time step is 0.06 fs, and the simulation stops at $t$ = 30 ps.

The purpose of this section is not to discuss the details of the processes leading to proton acceleration, but only to illustrate how plasma tailoring can improve the proton beam quality. The interplay between the different acceleration mechanisms and their dependence on the plasma tailoring scheme will be discussed in a forthcoming paper devoted to the results of these PIC simulations.

\begin{figure}
	\includegraphics[width=\columnwidth]{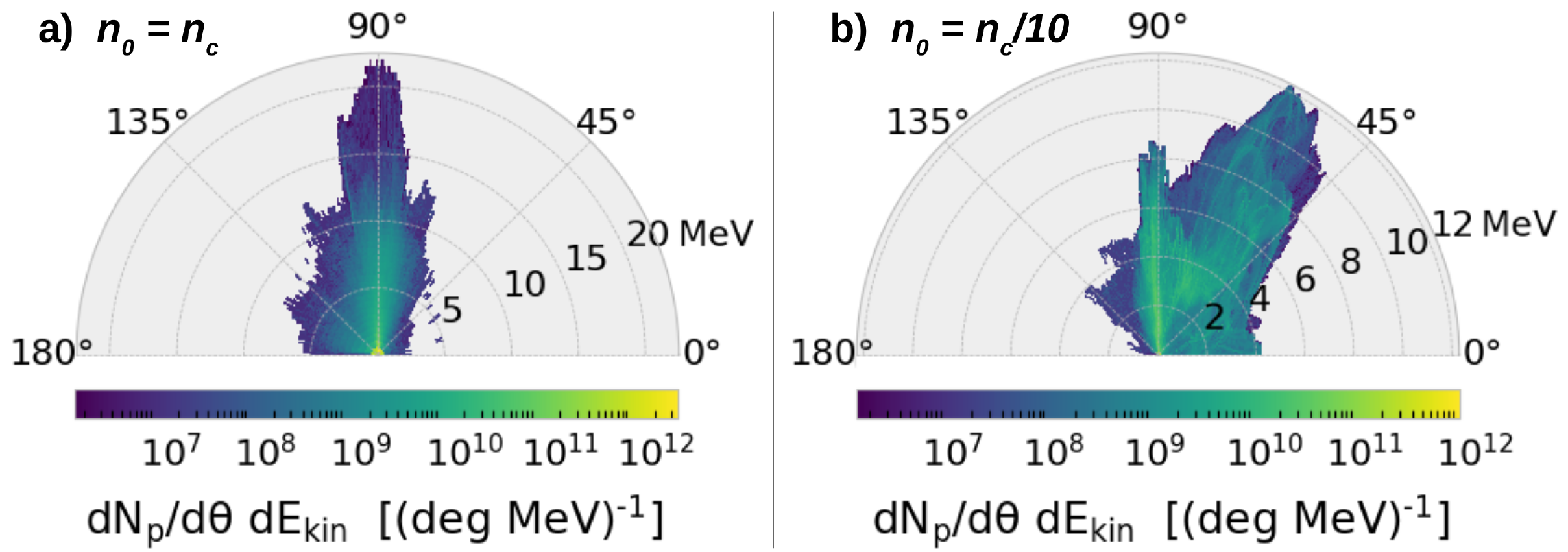}
	\caption{\label{Smilei_noshaping}\textit{Angular ($\theta$) and kinetic energy ($E_{kin}$) distribution of the accelerated protons \underline{without} plasma tailoring, for two initial plasma peak densities, a) $n_0$ = $n_c$, and b) $n_0$ = $n_c/10$. The density profile is the same than the one used in section \ref{plasma tailoring}: $n_{e0}(r) = n_0/[1+(r/r_0)^2]$ with $r_0$ = 70 $\mu$m, orange curve in Fig. \ref{1faisceau_2D}.}}
\end{figure}

The angular ($\theta$) and kinetic energy ($E_{kin}$) distribution of the accelerated protons in the gas jet, without any plasma tailoring, is presented in figure \ref{Smilei_noshaping}, for two initial plasma peak densities, $n_0$ = $n_c$, and $n_0$ = $n_c/10$. In the first case, the angular spectrum is strongly peaked at 90$^\circ$ from the laser axis. The energy spectrum is broad and extend up to $\sim$ 15-20 MeV. The number of protons accelerated along the laser axis is low, and their maximum energy is below 2 MeV. Without plasma tailoring, due to the high density in the wings of the gas jet, the growth rate of the filamentation instability is large and breaks the laser beam before it could reach the peak of the density profile. The protons are mainly accelerated at the beginning of the laser propagation. For example, the 15-20 MeV protons emitted at 90$^\circ$ are produced $\sim$ 350 $\mu$m before the laser focus (and the density peak).

Fig. \ref{Smilei_noshaping}-b) shows that when the plasma density is reduced by a factor 10, a second component appears in the angular distribution, at $\sim$ 70$^\circ$, and more protons are accelerated between 10 and 20 MeV. More protons are also accelerated along the laser axis, with a maximum energy that increases to $\sim$ 4 MeV, while the backward acceleration ($\theta > 90^{\circ}$) is almost suppressed. A 10-fold reduction of the plasma density leads to less filamentation of the laser, protons are accelerated all along its propagation, but still mainly in the transverse direction. The peak density is too low and the density gradient is too smooth to allow the generation of a longitudinal electrostatic shock and an efficient acceleration.

\begin{figure}
	\includegraphics[width=\columnwidth]{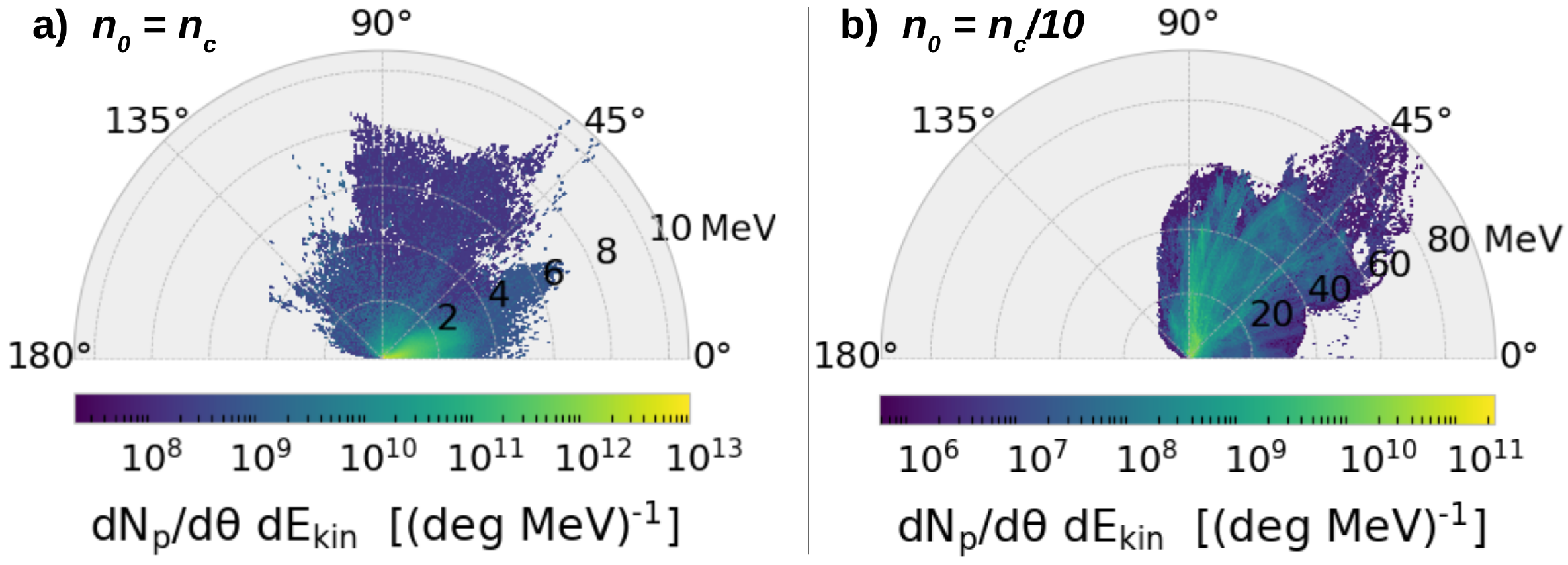}
	\caption{\label{Smilei_shaping_1beam}\textit{Angular and kinetic energy distribution of the accelerated protons obtained from a plasma \underline{tailored at the entrance side} of the ps-laser, for two initial plasma peak densities, a) $n_0$ = $n_c$ , and b) $n_0$ = $n_c/10$.}}
\end{figure}

The distribution of the accelerated protons obtained from a plasma tailored at the entrance side of the ps-laser is presented in figure \ref{Smilei_shaping_1beam}, for two initial plasma densities, $n_0$ = $n_c$, and $n_0$ = $n_c/10$. The density profiles are those obtained from TROLL simulations in the case $\Delta x$ = 250 $\mu$m and for the time where the BW is at the jet center (similar, for $n_0$ = $n_c$, to the purple curve in Fig. \ref{1faisceau_coupes}). Compared to the case without tailoring, the transverse acceleration (90 $^\circ$) is reduced in benefit of forward acceleration. For $n_0$ = $n_c$ (Fig. \ref{Smilei_shaping_1beam}-a)), the maximum energy at (0$^\circ$) increases from 0.4 MeV (no tailoring) to $\sim$ 3.2 MeV. However, the global plasma density is too high and the acceleration process is not very efficient. For an initial density 10 times lower (Fig. \ref{Smilei_shaping_1beam}-b)) the result is drastically different: the quality/intensity of the laser beam is preserved up to the sharp and high density ($n_c/2$) plasma slab, significantly increasing the number of high energy protons and their maximum energies: up to $\sim$ 17 MeV at 0$^\circ$, and more than 40 MeV at 45$^\circ$. As in the case without tailoring, reducing the density also suppresses the component in the backward direction (135$^\circ$).

\begin{figure}
	\includegraphics[width=\columnwidth]{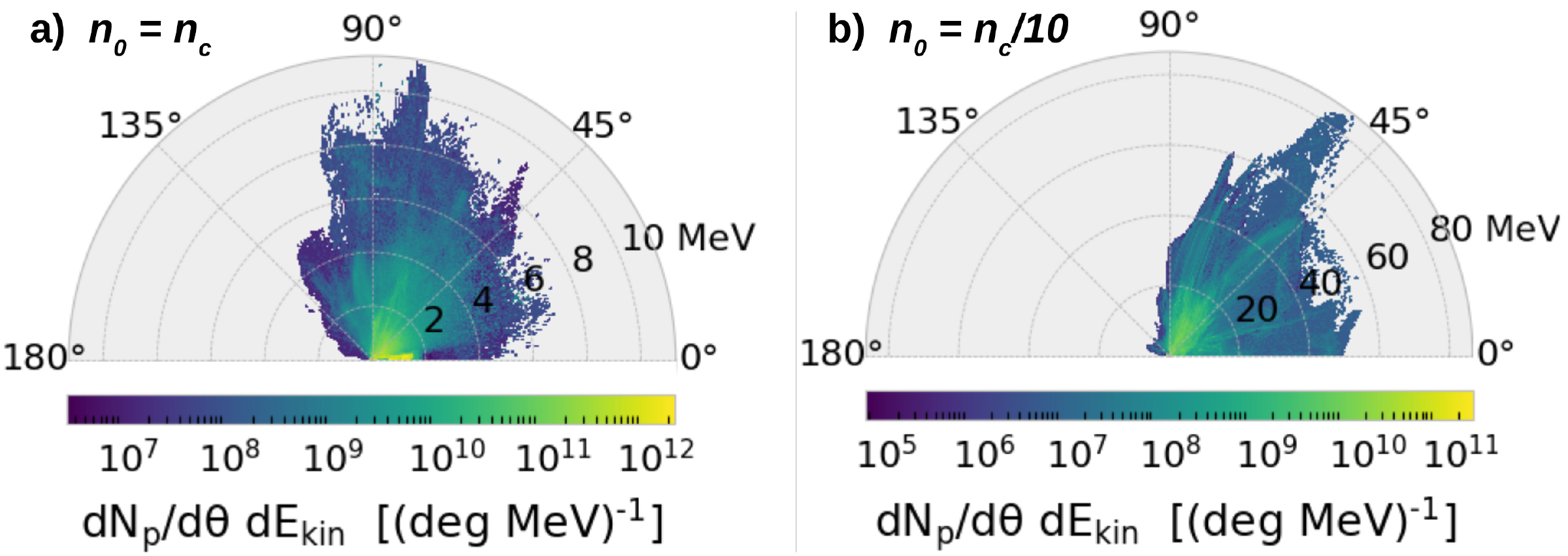}
	\caption{\label{Smilei_shaping_2beams}\textit{Angular and kinetic energy distribution of the accelerated protons obtained from a plasma \underline{tailored at the entrance and the exit sides} of the ps-laser, for two initial plasma peak densities, a) $n_0$ = $n_c$, and b) $n_0$ = $n_c/10$.}}
\end{figure}

The distribution of the accelerated protons obtained from a plasma tailored at both the entrance and the exit sides of the ps-laser is presented in figure \ref{Smilei_shaping_2beams}, for two initial plasma peak densities, $n_0$ = $n_c$, and $n_0$ = $n_c/10$. The density profile of the case $n_0$ = $n_c$ is the one presented in Fig. \ref{Coupes_2faisceaux_fcn_distance_laser} for $\Delta x$ = 250 $\mu$m - $t$ = 1.6 ns - blue curve. In that case, the peak density is too high ($\sim$ 17 $n_c$) and in addition, despite the tailoring, the plasma density in front of the sharp gradient is still large (0.03-0.1 $n_c$), the laser intensity profile is strongly perturbed, and the acceleration process weakly efficient, leading mainly to transverse acceleration with energies below 12 MeV.

At a lower initial density, $n_0$ = $n_c/10$ (Fig. \ref{Smilei_shaping_2beams}-b)), as observed for the single-side tailoring, the laser can propagate up to the density peak (2$n_c$), resulting in a much more efficient acceleration. The maximum energy is $\sim$ similar to the single-side tailoring case, but the number of protons accelerated to high energies is larger, together with a much weaker 90$^{\circ}$ component, and a more efficient acceleration along the laser axis, in terms of number of protons as well as of maximum energy: up to $\sim$ 50 MeV (3 times more).

\begin{figure}
	\includegraphics[width=\columnwidth]{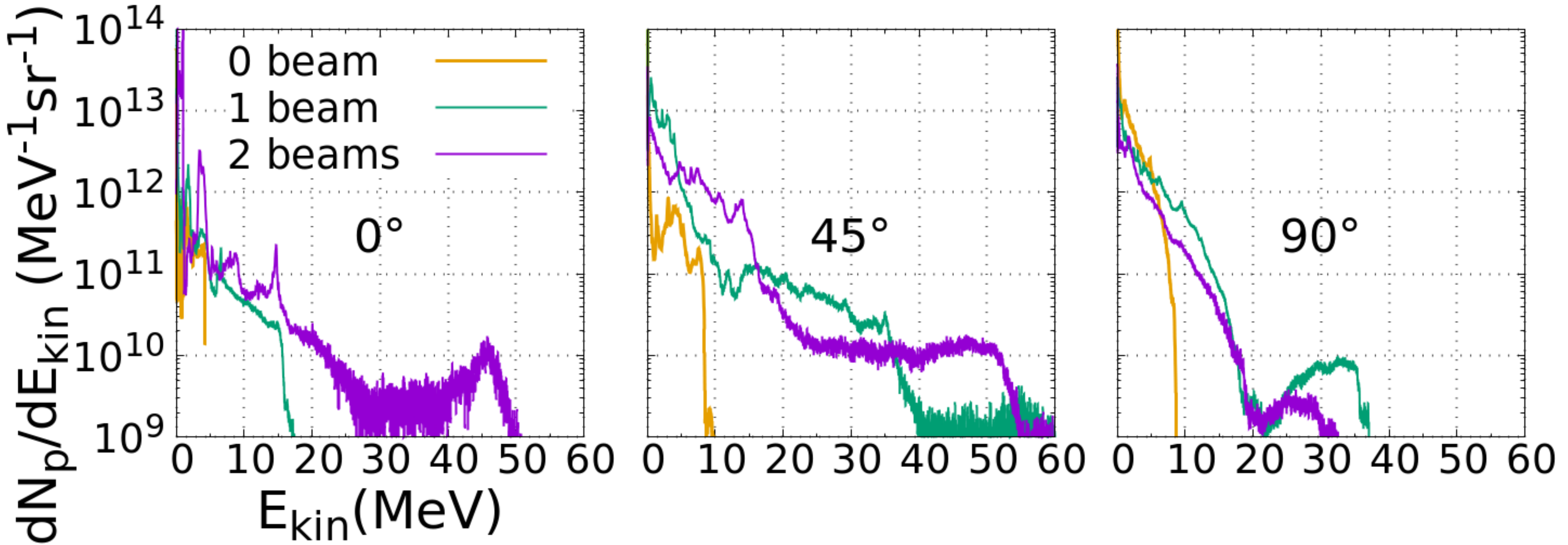}
	\caption{\label{Spectres_energie}\textit{Energy spectra at $\theta$ = 0$^\circ$, 45$^\circ$ and 90$^\circ$ in the case $n_{0}$ = $n_c/10$, and for the three types of plasma tailoring: none (0 beam), entrance side (1 beam), both sides (2 beams). Obtained by integrating the distribution $d^2N_p/d\theta dE_{kin}$ of Fig. \ref{Smilei_noshaping}-b), \ref{Smilei_shaping_1beam}-b) and \ref{Smilei_shaping_2beams}-b) in a $\pm$ 5$^\circ$ cone centered on the angle of interest.}}
\end{figure}

Figure \ref{Spectres_energie} presents the energy spectra $dN_p/dE_{kin}$ at $\theta$ = 0$^\circ$, 45$^\circ$ and 90$^\circ$ in the case $n_{0}$ = $n_c/10$, and for the three types of plasma tailoring: none, entrance side, both sides. Even if tailoring only the entrance side of the plasma can lead to high energy protons, tailoring both sides leads to a larger number of high energy protons and a more collimated beam. At 0$^\circ$ protons reach up to 50 MeV with a double-side tailoring, while tailoring only the entrance side only gives 17 MeV. Let us also not that in the case of double-side tailoring, the spectrum at 0$^\circ$ presents several narrow peaks (at $E_{kin}$ $\sim$ 4, 8, 15, 46 MeV), which could indicate that several acceleration mechanism could operate at the same time.

By integrating on the energy the spectra in Fig. \ref{Spectres_energie}, one can calculate the number of protons $N_p$ accelerated above a defined energy $E_s$. This is reported in Table \ref{Table_proton_numbers} as a function of $E_s$. It shows that tailoring both sides of the plasma density profile significantly increases the number of high energy protons, in particular along the laser axis, with for example $N_p = 2.7 \times 10^{11}$ protons above 15 MeV at 0$^\circ$, more than a factor 10 compared to the case with only the entrance side tailored. The ratio $N_p(0)/N_p(45)$ is also significantly increased: from 1.6$\%$ with 1 beam to 29 $\%$ with 2 beams.

The purpose of these PIC simulations is to illustrate that tailoring the plasma on both sides allows to increase the number of high energy protons as well as shifting their angular distribution toward the laser axis. This last effect is very important since the angular distribution is symmetric around the laser axis (conical emission), so that in order to collect as much protons as possible one needs to reduce as much as possible the angle $\theta$ of this cone. The results presented in section \ref{plasma tailoring} show that several parameters governs the plasma profile and can be adjusted to optimize the proton beam, like for example the timing between the ns tailoring-beams and the ps pulse in order to adjust the length of the high density region (Fig. \ref{Coupes_1beam_late_time} and \ref{Coupes_temps_long_2beams}), or the distance $\Delta x$ of the ns-beam to optimize the contrast of the density gradient.

\begin{table}
	\begin{center}
		\begin{tabular}{|c||c||c|c|c|c|} \hline
			$E_s(MeV)$&\# of beams & $N_p(0^{\circ}$) & $N_p(45^{\circ}$) & $N_p(90^{\circ}$)\\
			\hline \hline
			$\ge 1$&0&8&23&220\\
			\cline{2-5}&1&23&278&172\\
			\cline{2-5}&2&43&200&108\\
			\hline \hline
			$\ge$ 5&0&0&6&19\\
			\cline{2-5}&1&5&39&62\\
			\cline{2-5}&2&12.4&109&26\\
			\hline \hline
			$\ge$ 10&0&0&0&0\\
			\cline{2-5}&1&1.8&17&14.4\\
			\cline{2-5}&2&6.3&40&5.6\\
			\hline \hline
			$\ge$ 15&0&0&0&0\\
			\cline{2-5}&1&0.2&12.6&1.8\\
			\cline{2-5}&2&2.7&9.3&0.9\\
			\hline
		\end{tabular}
	\end{center}
	\caption{\textit{Number $N_p$ of protons accelerated above $E_s$, as a function of the number of beams tailoring the plasma, and emitted at 0$^\circ$, 45$^\circ$ or 90$^\circ$. $N_p$ is obtained from the integration of the spectra of Fig. \ref{Spectres_energie} on $E_{kin} = [E_s,\infty]$, and is expressed in $10^{11}$ MeV$^{-1}$.sr$^{-1}$.}}
	\label{Table_proton_numbers}
\end{table}

\section{Conclusions}
Through multi-dimensional single-fluid hydrodynamic as well as Fokker-Planck simulations, we have studied a new scheme of plasma tailoring by lasers, that offers attractive capabilities and can easily be implemented, since it is based on a commonly used high-density narrow hydrogen gas jet coupled with widespread nanosecond laser pulses. Compared to other schemes, the use of lasers propagating along the sides of the jet (and not toward its center) allows tailoring of both sides of the jet while avoiding a counter-propagating geometry, which is often risky for the laser chain. It is thus possible to create a thin plasma slab with a sharp gradient of high contrast on both sides. 

In the scenario of a single-side tailoring, compared to methods using a laser focused at the jet center, that generate a spherical diverging BW, the present scheme leads to a BW with a converging bow shape that compresses the plasma from the side toward the center, which creates a plasma slab of constant density but with a thickness that increases with time, offering a possibility to adjust the plasma width.

By tailoring both sides of the jet, the collision of the BWs enables to increase (compared to a single-side tailoring) the peak density by a factor 2-3, while the stagnation regime reached at the collision point slows down the plasma expansion, offering some adjustment on the width of the plasma slab while preserving a high density.

In a range of at least one order of magnitude, the shape of the final profile is almost independent on the initial plasma density, so that the peak density can simply be adjusted by the backing pressure of the gas jet. This device is debris free and can answer the challenge of providing targets for high repetition rate laser facilities.

Since the laser is focused in a low density part, the efficiency of the energy deposition is low ($\le$ few $\%$), and the BW generation requires more energy (few hundreds of mJ) than schemes based on a laser focused at the jet center or on a solid target placed at the edge of the jet. However, once the BW is excited, its expansion weakly depends on the laser energy, so that the position and arrival time of the high-density peak are very robust to laser fluctuations, including also the focal spot quality and pointing stability. A possible way to reduce the required laser energy could be the use of a jet of clusters\cite{Ditmire,Smith,Marocchino}. 
Using 2D-PIC simulations, the benefit that such a target could bring has been evaluated on the acceleration of high energy protons by sub-ps high intensity laser. Compared to steepening only the entrance side of the gas jet, tailoring also the backside not only increases significantly the total number of high energy protons and their maximum energy, but also shifts their angular distribution toward the laser axis, which is crucial to produce collimated proton beams usable for applications. Since the proton acceleration mostly takes place near the plasma front, reducing the amount of "unused" plasma that the proton beam has to cross before reaching vacuum could also be of interest in order to avoid beam instabilities that could deteriorate the beam quality.

\begin{acknowledgments}
We acknowledge the help of Mickael Grech on the code Smilei. The PIC simulation work was granted access to HPC resources of TGCC under the allocation A0010506129 made by GENCI and under the allocation 2017174175 made by PRACE.

The data that support the findings of this study are available from the corresponding author upon reasonable request.
\end{acknowledgments}

\bibliography{references}

\end{document}